# Mode interaction aided excitation of dark solitons in microresonators constructed of normal dispersion waveguides

Xiaoxiao Xue[1], Yi Xuan[1,2], Yang Liu[1], Pei-Hsun Wang[1], Steven Chen[1], Jian Wang[1,2], Dan E. Leaird[1], Minghao Qi[1,2], and Andrew M. Weiner[1,2*]

[1]*School of Electrical and Computer Engineering, Purdue University, 465 Northwestern Avenue, West Lafayette, Indiana 47907-2035, USA*

[2]*Birck Nanotechnology Center, Purdue University, 1205 West State Street, West Lafayette, Indiana 47907, USA*

*amw@purdue.edu

## Abstract

Kerr frequency combs from microresonators are now extensively investigated as a potentially portable technology for a variety of applications. Most studies employ anomalous dispersion microresonators that support modulational instability for comb initiation, and mode-locking transitions resulting in coherent bright soliton-like pulse generation have been reported. However, some experiments show comb generation in normal dispersion microresonators; simulations suggest the formation of dark pulse temporal profiles. Excitation of dark pulse solutions is difficult due to the lack of modulational instability in the effective blue-detuned pumping region; an excitation pathway has been demonstrated neither in experiment nor in simulation. Here we report experiments in which dark pulse combs are formed by mode-interaction-aided excitation; for the first time, a mode-locking transition is observed in the normal dispersion regime. The excitation pathway proposed is also supported by simulations.

Microresonator-based optical frequency combs, also termed Kerr combs, are generated through conversion of a single pump frequency to a broadband frequency comb inside a high-quality-factor (*Q*) microresonator via the third-order Kerr nonlinearity [1]-[10]. The advantages of Kerr combs include very compact size, high repetition rate, and capability of generating ultra-broad combs.

The dynamics of Kerr comb generation have attracted intense investigations since the first demonstration of the method [11]-[28]. It has been found that Kerr combs are not always coherent [11]-[12] and may be characterized by high intensity noise [13]-[14]; furthermore, lack of coherence and high intensity noise are generally correlated. Experiments have revealed transitions from low coherence, high noise states to highly coherent mode-locked states accompanied by a sudden drop in the comb noise [14]-[18]. It has been found in simulations and experiments that the mode locking of broadband Kerr combs is usually related to soliton formation in the cavity [15], [17]-[28]. These dissipative cavity solitons are localized structures stabilized by a balance between Kerr nonlinearity and dispersion. In time domain they exist as bright or dark pulses, depending on whether the cavity dispersion is anomalous or normal, respectively. Bright microresonator solitons in the anomalous dispersion region have been observed in experiments and well studied through simulations [15], [17]-[27]. Reference [17] reported a method of tuning the pump laser frequency to an effectively red-detuned regime (pump laser wavelength longer than resonant wavelength) which is typically difficult to achieve due to thermal instability [29]. Mode-locking transitions yielding bright solitons were observed after passage through a broadband chaotic state [17]. In contrast, although dark solitons have been predicted in normal dispersion microresonators in theory and simulation [27]-[28], investigating dark solitons experimentally is extremely difficult and no time-domain characterization has ever been reported. The first challenge is spontaneous excitation of dark solitons. Although modulational instability (which is required for the growth of primary comb lines from a continuous-wave field with noise) is possible in normal dispersion resonators with the proper laser-resonance detuning [30]-[32], it requires the pump laser frequency be in the red-detuned region which is generally unstable due to thermal instability [29] and modulational instability induced intracavity power switching [33]. Moreover, it is difficult to directly characterize dark solitons with conventional ultrafast techniques such as frequency resolved optical gating (FROG) or second order autocorrelation. Although experimental comb spectra were ascribed to dark pulse action in [34] and [35], no corroborating measurements of spectral phase or time domain profiles were reported, and an excitation pathway for dark soliton formation was not proposed. In this work, we show experimentally that dark solitons can be excited with the aid of mode interactions in microresonators which are constructed of normal dispersion waveguides. For the first time, mode-locking transitions related to dark soliton formation are observed in the normal dispersion regime, and we demonstrate the time-domain characterization of dark solitons by first converting them to bright pulses via line-by-line pulse shaping [11] and then performing autocorrelation or crosscorrelation measurements. The mode interaction aided excitation of dark solitons appears to occur through a deterministic pathway, in sharp contrast to the situation for bright solitons, where the number of solitons generated is stochastic due to the pathway through a chaotic state [17], [22].

The nonlinear microresonator we employed is silicon nitride (SiN) microring which can be fabricated by using the CMOS-compatible technique (see Methods) [3]-[4], [11]. In the first example, an Au microheater is also integrated together with the SiN microring (microscope image shown in Fig. 1a) which can be used to shift the resonance frequencies via the thermal-optic effect (see Methods). The microheater-based thermal tuning technique gives us an alternative ultra-stable way of controlling the pump-resonance detuning in addition to the conventional method of only changing the pump laser frequency. The free spectral range (FSR) of the mode used for comb generation is 231.3 GHz and the loaded resonance width is 250 MHz (corresponding to a loaded *Q* of $7.7\times10^{5}$). The measured group-velocity dispersion (GVD) is $\beta_2 = 190.7 \pm 8.4\ \mathrm{ps}^2/\mathrm{km}$ [36]. Figure 1b shows the deviation of the resonance frequencies $\omega_\mu = \omega_0 + D_1\mu + \frac{1}{2}D_2\mu^2 + \ldots$ from an equidistant frequency grid defined by $\omega_0 + D_1\mu$ [5], [14], [17], where $\omega_0$ is the resonance pumped and $\mu$ the relative mode number. $D_1$ is the FSR at $\omega_0$; $D_2 = -c/n_0 D_1^2 \beta_2 \approx -2\pi 10$ MHz the approximate FSR change for three adjacent resonances, where $c$ is the light speed in vacuum and $n_0$ the refractive index. The resonance deviation shows as a negative quadratic function of the relative mode number, which is evidence of a strong normal dispersion.

For comb generation (setup shown in Fig. 1c), we tuned the microresonator by changing the voltage applied to the microheater to match the resonance wavelength with the pump laser wavelength. The pump laser wavelength was fixed at 1549.3 nm which was initially to the red of the resonance used (around 1548.4 nm) when the heater power was zero; and the pump power was around 1.7 W (off-chip power, the coupling loss per facet is around 3 dB). The heater voltage was first increased to 20 V (the corresponding resonance was red-shifted to around 1549.5 nm, which was now to the red of the pump), and then slowly reduced to shift the resonance back toward the pump. When the heater voltage was 7.95 V, there were two primary comb lines far away (54-FSR) from the pump, and also some secondary comb lines with 1-FSR spacing (Fig. 1d I). The comb at this stage showed broadband intensity noise accompanied by some peaks (Fig. 1e I). When the heater voltage was further reduced to 7.73 V, a sudden transition to a broadband comb was observed (Fig. 1d II), still with intensity noise but now appearing as a narrow noise peak around 700 MHz (Fig. 1e II). The narrow RF peak signifies quasi-periodic slow evolution of the comb, often termed a “breather” [24], [37]. When the heater voltage was further reduced to 7.34 V, the comb spectrum remained similar (Fig. 1d III) while the intensity noise dropped below the background noise of the RF spectrum analyzer (Fig. 1e III). In experiments, we found that a similar low-noise transition could be achieved alternatively by slightly reducing the pump power with the pump wavelength and heater voltage fixed. The low-noise transition behavior and the mode-locked state were also verified by measuring the beat note of the comb line with a narrow-linewidth reference laser (see Supplementary Information Section 1).

Our microresonator-tuning method is equivalent to the traditional method of tuning the pump laser wavelength from the blue side to stably approach the resonance which is red shifted by the thermo-optic effect and Kerr effect [17]. This leads to the intracavity pump field staying on the upper branch of the bistability curve where modulational instability is generally absent [30]-[32] (see Supplementary Information Section 3). We believe the initial comb lines shown in Fig. 1d I were formed due to interaction of different family modes [18], [38]-[40]. In overmoded microresonators, mode coupling may occur between different mode families around mode crossing positions. The resonant frequency of each mode is shifted, which may equivalently be viewed as an additional per round trip phase shift for each of the corresponding comb frequency components. The phase relationship between the pump mode and the two sideband modes, one of which is coupled to the other transverse mode, may be affected such that an equivalent anomalous dispersion is achieved [38]-[40]. Modulational instability thus occurs and generates some initial comb lines.

To provide evidence for the mode interaction, we pumped different resonances belonging to the same transverse mode family as that pumped in Fig. 1d. The pump power was reduced to 0.6 W so that only two primary comb lines were generated in the optical spectrum analyzer range (Fig. 1f). The pump was shifted by a total of 2.76 THz (12 FSRs), while the long wavelength sideband varied by no more than $\pm 15$ GHz. The short wavelength sideband varied at twice the rate of pump tuning, for a total variation of 5.52 THz. The long wavelength sideband was always anchored at the same position which is a signature of mode-interaction-aided comb generation [39].

The thermal tuning rates of different mode families are different. For the SiN microring shown in Fig. 1a, another transverse mode can be observed besides the one used for comb generation. When the mode pumped for comb generation is tuned by 1 nm, the two transverse modes have a differential shift estimated around 37 pm, which is much larger than the resonance width (2 pm FWHM) involved in comb generation. The strength of mode interaction also depends on the resonance wavelength, i.e. the interaction may be strong around one wavelength but nearly absent around another [39]. These facts imply that the details of the mode interaction can be changed when we thermally tune the microresonator using the microheater. To further investigate the relationship between mode interaction and the broadband comb in Fig. 1d, we shifted the entire mode-locked comb spectrum after transition by tuning the pump laser wavelength and the microresonator in tandem (i.e., tuning them in turn in a small step <0.01 nm). Figure 2a shows the tunable comb spectra measured with the pump wavelength tuned over a total range of 1549 nm to 1550.4 nm. Except for lines near the mode interaction areas at the edges of the spectral envelope, the amplitude of each comb line remained nearly constant. A low noise state could be achieved for all the combs by optimizing the laser-microresonator detuning. It can be observed from Fig. 2a that the comb cluster around 1660 nm which was mainly attributed to mode interaction disappeared when the pump laser wavelength was tuned below 1549.1 nm or above 1549.6 nm; thus the comb spectrum became more symmetric. The results suggest that the mode interaction was changed when the microresonator was thermally tuned to match the pump wavelength in these ranges. Our experiments also showed that the broadband comb could not be generated with the pump laser fixed in approximately the same range below 1549.1 nm or above 1549.6 nm where the mode interaction induced comb cluster was absent. The broadband comb was preferably generated with the pump fixed in the range above 1549.1 nm and below 1549.6 nm, but then might be tuned outside of this range. The experimental evidence shows that the broadband comb, even though excited by mode interaction induced initial combs, was not governed by the mode interaction. There exists another mechanism responsible for the stable broadband comb which in the following is recognized as dark solitons.

We used a pulse shaper in the lightwave C band to perform line-by-line shaping of the mode-locked frequency comb. Twenty one comb lines fell in the frequency range of the pulse shaper. The power spectrum was shaped to a smooth Gaussian profile (Fig. 2b). The phase of each comb line was then compensated to form a transform-limited pulse train (see Methods) [11]. All the combs at different pump wavelengths shown in Fig. 2a could be cleanly compressed, with high contrast, to a duration of ~318 fs (autocorrelation width, corresponding to ~220 fs pulse width) (Fig. 2c), which indicates that high coherence was maintained. The compressed autocorrelations measured for 15 different combs shifted by 0.1 nm wavelength increments are overlapped and are essentially indistinguishable. The retrieved comb phase curves are shown in Fig. 2d. A small variation of the phase curve can be observed during the comb tuning process, which coincides with the change in the optical spectrum of the comb shown in Fig. 2a. We attribute this to the change of mode interaction. Nevertheless, the phase curves remain similar. The repeatability of this state of the comb is extremely good. In Fig. 2d, there is also one comb phase curve which was measured on a different day with the pump wavelength 1549.3 nm. A result very similar to the other curves was obtained. The ability to maintain a nearly unchanged low noise state, even while tuning nearly a full FSR and experiencing changes in the mode interactions that initiate the comb, signifies the robust character of the discovered mode-locked operating regime.

By using the spectral phase information (Fig. 2d) retrieved via line-by-line pulse shaping and the power spectrum measured without amplitude shaping, we can reconstruct the time-domain waveform in the microresonator. To do so, the amplitude and phase of the pump line measured at the through port are first corrected to represent the component from coupling out of the microring cavity (see Supplementary Information Section 4 both for discussion of the analysis procedure and for corroborating experiments obtained for another resonator fabricated with a drop port). The reconstructed waveforms at different pump wavelengths are shown in Figs. 2e & 2f which are square dark pulses with chirped edges and chirped ripples at the bottom. The dark solitons show complex structure, in sharp contrast to isolated pulse bright solitons in the anomalous region (see Supplementary Information Section 5 for results measured with a larger pulse shaping range).

The mode interaction induced initial comb show in Fig. 1d I consists of primary comb lines and 1-FSR spacing secondary combs. In experiments, we found that primary comb lines may also excite dark solitons directly. Figure 3 shows the results of another microring example fabricated with both through port and drop port [41]. The advantage of a drop port is that the intracavity field may be observed without the complication of a strong superimposed pump field. The radius of this microring is 100 μm and the loaded $Q$ is $8.6\times10^5$. The waveguide constructing the microring (microscope image shown in Fig. 3a) has nominally the same dimensions as the one in Fig. 1a, and the measured dispersion ($\beta_2 = 186.9 \pm 7.3\ \mathrm{ps^2/km}$) matches within experimental error (Fig. 3b). Here, strong mode interactions occur for mode 1 ($\mu$=1) and mode 2 ($\mu$=2), which can be concluded from the corresponding resonance jumps in Fig. 3b. For comb generation, the pump laser frequency was tuned slowly into the resonance from the blue side. The pump power was around 1 W. In two tests, mode 1 and mode 0 were pumped, generating 1-FSR and 2-FSR combs respectively. Figures 3c and 3f show the optical power measured at the drop port when the pump laser is scanned across the resonance from the blue side (scanning speed 0.5 nm/s). Similar to the case of bright soliton formation in the anomalous dispersion region [17], power drop steps which indicate transition behavior can be observed here. It has been shown that the power drop steps related with bright soliton formation randomly changes from scan to scan which implies a stochastic transition pathway [17]. In contrast, in Figs. 3c and 3f we actually overlay twenty measurements under repeated laser scans; the traces are virtually identical, giving evidence for a deterministic pathway toward mode-locking in the normal dispersion region. The comb spectra and intensity noise at different stages are shown in Figs. 3d, 3e (for the 1-FSR comb) and 3g, 3h (for the 2-FSR comb) respectively. A low-noise mode-locking transition was observed in both cases. The transition behavior was also verified by measuring the beat note of a comb line with a narrow-linewidth reference laser (see Supplementary Information Section 1).

The 1-FSR and 2-FSR mode-locked combs shown in Figs. 3d III and 3g III have similar envelope, suggesting that the localized feature of time-domain waveform does not depend on the repeat period. Since the pumping wavelength is near the edge of our pulse shaper's passband, we used a different method incorporating cross-correlation measurement to investigate the intracavity time-domain waveform (Fig. 4a). Part of the comb power from the drop port was compressed to a short transform-limited pulse train by using a pulse shaper, and then used as a sampling signal to test the waveform from the drop port. Figures 4b and 4c show the measured cross-correlation results for the 1-FSR and 2-FSR combs respectively. In both cases, the cross-correlation shows a series of dark pulses with approximately the same shape. The width of the dark pulse is ~700 fs. The numerical simulation results based on the estimated experimental parameters are shown in Figs. 4d and 4e (see Fig. 5 in the following for the simulation details). The simulated dark pulse cross-correlation width is ~800 fs which is close to the experimental observation.

When the pump laser is tuned into the resonance from the blue side, the resonance is red shifted due to thermo-optic and Kerr nonlinearities. The strength of mode interaction may be changed in this process. Figure 3b also shows the measured resonance frequencies for a pumped cavity. The laser was pumping mode −7 with 1 W. The red detuning of the pump laser with respect to the cold-cavity resonance was around 0.5 nm. A second scanning probe laser was used to measure the microring transmission at the drop port. No combs were generated in this measurement. The resonance jumps of mode 1 and mode 2 are changed in comparison to the cold cavity, which shows clear evidence of the mode interaction change in a pumped cavity. However, in the case of comb generation, the details of mode interaction change are very difficult to measure in experiments by scanning the transmission, due to the interference of comb lines and the mixture of thermal effect and Kerr effect.

To further model the dark solitons, numerical simulations are performed by using the standard Lugiato-Lefever (L-L) equation [19], [30], modified to include mode interaction (see Methods and Supplementary Information Section 2). Figure 5 shows the simulation results for the microring in Fig. 3a. The intracavity field at the start of simulation is the steady-state continuous-wave solution on the upper branch of the bistability curve plus weak noise. Figures 5a shows the evolution of optical spectrum versus the slow time when mode 1 is pumped. A 1-FSR comb is generated. The transient images of the comb spectrum and the time-domain waveform are shown in Figs. 5b and 5c respectively. The detuning of the pump laser with respect to the cold-cavity resonance is increased at 45 ns after the mode interaction induced initial comb lines grow up. The comb then transitions to a breather state which changes periodically over time. This behavior is similar to that of the experimental result shown in Fig. 3d I. The measured comb intensity noise contains several narrow peaks, which indicates quasi-periodic changing of the comb state. In the simulation, the breather comb transitions to a stable state after the detuning is increased further at 105 ns. The time-domain waveform shows one dark pulse per roundtrip. Figures 5d, 5e, and 5f show the simulation results for the 2-FSR comb when mode 0 is pumped. After the detuning is increased at 200 ns, the comb transitions to a stable mode-locked state which is close to the experiment result shown in Fig. 3g III. The time-domain waveform show two dark pulses per roundtrip. The excitation behavior of dark solitons revealed by simulations is similar to our experimental observations. The deviations between simulation and experiment are partially attributed to dynamic change of mode interactions during the pump tuning process, which are difficult to measure and hence difficult to capture exactly in the simulation. Similar numerical simulations, showing good agreement with our measurements for the microring of Fig. 1a, are shown in Supplementary Information Section 5.

Further simulation results reveal that the generation of stable dark solitons and breathers is related to interaction of fronts which connect the two steady-state solutions of the continuous-wave bistability curve (see Supplementary Information Section 6). The theories of moving fronts (which are also named "switching waves") and soliton formation in a driven nonlinear cavity can be found in Rosanov's work as well as some review articles [42]-[44]. In our experiments and simulations, the fronts are formed due to the modulational instability enabled by mode interaction. This process is usually accompanied by growing of the primary comb lines and wave breaking [45]. When two fronts are close to each other, they may be trapped by each other to generate stable localized structures [42]-[44]. Our experimental and simulation results show that the physics of switching waves and dark solitons is highly relevant for the generation of broadband mode-locked microresonator combs especially in the normal dispersion regime.

In summary, we have demonstrated dark soliton formation as well as mode-locking transitions in normal dispersion microresonators. Such dark solitons in the normal dispersion regime have very different nature from bright solitons in the anomalous dispersion regime [46]. Mode-interaction-aided excitation of dark solitons is reported. Besides high scientific interest, microresonator dark solitons also have their advantages in practical applications. As most nonlinear materials have normal dispersion, tailoring of the microresonator geometry is generally required to get overall anomalous dispersion [5], [47]-[48]. Although such waveguide engineering works well around the telecom band, the ability to mode-lock in the normal dispersion regime, as reported here, increases freedom in microresonator design and may make it possible to generate Kerr combs in an extended wavelength range. This might prove especially important in the visible, where material dispersion is likely to dominate. Furthermore, in the process of dark soliton formation, the steady-state intracavity power (i.e., the background of the dark solitons) stays on the upper branch. The intracavity energy does not change too much after dark solitons are formed; thus there is no severe thermal instability problem. The pump laser frequency or the microresonator does not need to be tuned at a carefully selected speed to overcome the thermal instability issue as needed for bright soliton formation in the anomalous dispersion region [17]. This may allow reduced system complexity. Another advantage of dark soliton combs demonstrated by our experiments is the excellent repeatability. Similar transition behavior can be achieved each time the pump laser is tuned into the resonance with the same power. The good repeatability of dark solitons is related to the excitation pathway which is regulated by mode interaction. In comparison, because bright solitons are excited by the broadband chaotic state before the mode-locking transition, the transition exhibits a stochastic behavior, meaning that different numbers of soliton pulses may be generated under essentially identical experimental conditions [17].

## Methods

**Device fabrication.** An under cladding layer of 3 um thermal oxide is grown on a silicon wafer in an oxidation tube at 1100 °C. Using LPCVD, a 550 nm SiN film is deposited at 800 °C on the oxidized wafer. A negative HSQ resist is used to pattern the waveguide and resonator via an EBL system at 100 kV. After developing in TMAH solution, the HSQ pattern is transferred to the SiN film using reactive ion etching. Then, a 3.5 um thick LTO film, which serves as an upper cladding, is deposited at 400 °C followed by an annealing step undertaken at 1100 °C in an $N_2$ atmosphere. Finally, Au(300 nm)/Cr(5 nm) is deposited on the upper cladding right above the resonator to form microheaters. The radius of the SiN microring is 100 μm. The cross-section dimension of the microring waveguide is 2 μm X 550 nm. The width of the bus waveguide is 1 μm. The coupling gap between the bus waveguide and the microring is 300 nm for the microring in Fig. 1a, and 500nm for the microring in Fig. 3a.

**Thermal tuning.** The resistance of the microheater shown in Fig. 1a is 291 ohm. By changing the voltage applied to the microheater (thus changing the heating power), the mode used for comb generation can be tuned with an efficiency of 0.82 nm/W. The thermal tuning efficiency of the second mode family, which was not pumped for comb generation, is 0.79 nm/W.

**Line-by-line comb shaping.** A commercial pulse shaper (Finisar WaveShaper 1000S) is used which has a spectral resolution of 10 GHz and a frequency setting resolution of 1 GHz. The method of line-by-line phase compensation to form a transform-limited pulse is the same as in [11]. The use of this method to determine the spectral phase of the as-generated waveform has been reported in a number of previous studies, and its validity has been confirmed by direct comparison to independent measurement methods [49]-[50].

**Autocorrelation and crosscorrelation.** An Erbium-doped fiber amplifier (EDFA) is used after the pulse shaper to compensate the link loss (not shown in Fig. 1c and Fig. 4a). In the autocorrelation measurement (Fig. 1c), a length of dispersion-compensating fiber (DCF) is used to roughly compensate the dispersion of the fiber link between the output port of the microring chip and the input ports of the correlator. The residual second-order dispersion (group velocity dispersion) as well as higher-order dispersion are measured by injecting ~180-fs pulses from a mode-locked fiber laser into the fiber link and are further compensated by programming the WaveShaper. In the crosscorrelation measurement (Fig. 4a), the fiber length between the output port of the microring chip and the input port 2 of correlator is much shorter than that in the autocorrelation measurement since there is no pulse shaper and EDFA in this path. The dispersion is then compensated by just using a short length of DCF.

**Numerical simulation.** Figure 5 shows the simulation results for the microring shown in Fig. 3a. The round-trip cavity loss $\alpha = 3.10\times10^{-3}$, and the bus waveguide coupling coefficient $\theta = 1.93\times10^{-3}$, are extracted by measuring the resonance width and the coupling condition (see Supplementary Information Section 4.4). The estimated nonlinear coefficient is

$$\gamma = \frac{n_2\omega_0}{cA_{eff}} \approx 0.89\ \mathrm{m^{-1}W^{-1}}$$

where the nonlinear refractive index $n_2 = 2.4\times10^{-19}\ \mathrm{m^2W^{-1}}$; the resonance frequency $\omega_0 = 2\pi\times1.95\times10^{14}$ Hz; $c$ is the light speed in vacuum; the effective mode area $A_{eff} \approx 1.1\times10^{-12}\ \mathrm{m^2}$. The initial intracavity field is the steady-state continuous-wave solution on the upper branch of the bistability curve plus weak noise (~ 1 pW/mode). The mode interaction is taken into account by applying additional phase shifts to mode 1 and mode 2 in the frequency domain step of the split-step Fourier routine [38]-[39]. The additional phase shift per round-trip $\Delta\phi$ and the corresponding resonance shift $\Delta f$ are related by $\Delta\phi = -2\pi\cdot\Delta f/FSR$. The resonance shifts of mode 1 and mode 2 in simulation are roughly of the same order as those obtained in measurements (see Fig. 3b). For the 1-FSR comb simulation, mode 1 is pumped with 0.4 W. The initial phase detuning of the pump laser with respect to the cold cavity is $\delta_0 = 1.2\times10^{-2}$ rad at the beginning of the simulation; and the mode interaction induced resonance shifts of mode 1 and mode 2 are $-130$ MHz and $230$ MHz respectively. After the initial comb lines are generated, the pump phase detuning is increased to $1.4\times10^{-2}$ rad at slow time 45 ns. To simulate the dynamic change of mode interaction with detuning, the resonance shifts of mode 1 and mode 2 are changed to $-100$ MHz and $100$ MHz respectively. The comb then transitions to a breather state which is similar to the experimental result in Fig. 3d I. The pump detuning is increased further to $1.85\times10^{-2}$ rad at slow time 105 ns; and the resonance shifts of mode 1 and mode 2 are changed to $-80$ MHz and $50$ MHz respectively. The comb transitions to a stable mode-locked state which is close to the experimental result in Fig. 3d III. For the 2-FSR comb simulation, mode 0 is pumped with 0.3 W. The initial pump detuning is $1.95\times10^{-2}$ rad; and the resonance shifts of mode 1 and mode 2 are $-130$ MHz and $230$ MHz respectively. The pump detuning is increased to $2.15\times10^{-2}$ rad at 200 ns; and the resonance shifts of mode 1 and mode 2 are changed to $-80$ MHz and $50$ MHz respectively. A stable mode-locked 2-FSR comb is obtained which is similar to the experimental result in Fig. 3g III. Note that the L-L equation generally uses an $\exp(-i\omega t)$ convention [19], [30], a convention which we adopt also in our simulations. However, for our experimental measurements we use the $\exp(i\omega t)$ convention prevalent in ultrafast optics [51]. Therefore, in comparing simulation with experiment, one must take into account the opposite sign conventions.

### Author contributions

X.X. led the experiments with assistance from Y.L., P.H.W., S.C., J.W., and D.E.L. X.X. analyzed the data and performed the numerical simulations. X.X and Y.X. designed the SiN microring layout with assistance from P.H.W. and J.W. Y.X. fabricated the microring. X.X. and A.M.W. wrote the manuscript. The project was organized and coordinated by A.M.W. and M.Q.

### Acknowledgements

This work was supported in part by the National Science Foundation under grants ECCS-1102110 and ECCS-1126314, by the Air Force Office of Scientific Research under grant FA9550-12-1-0236, and by the DARPA PULSE program through grant W31P40-13-1-0018 from AMRDEC. We thank Curtis Menyuk and Giuseppe Daguanno for helpful discussion and comments.

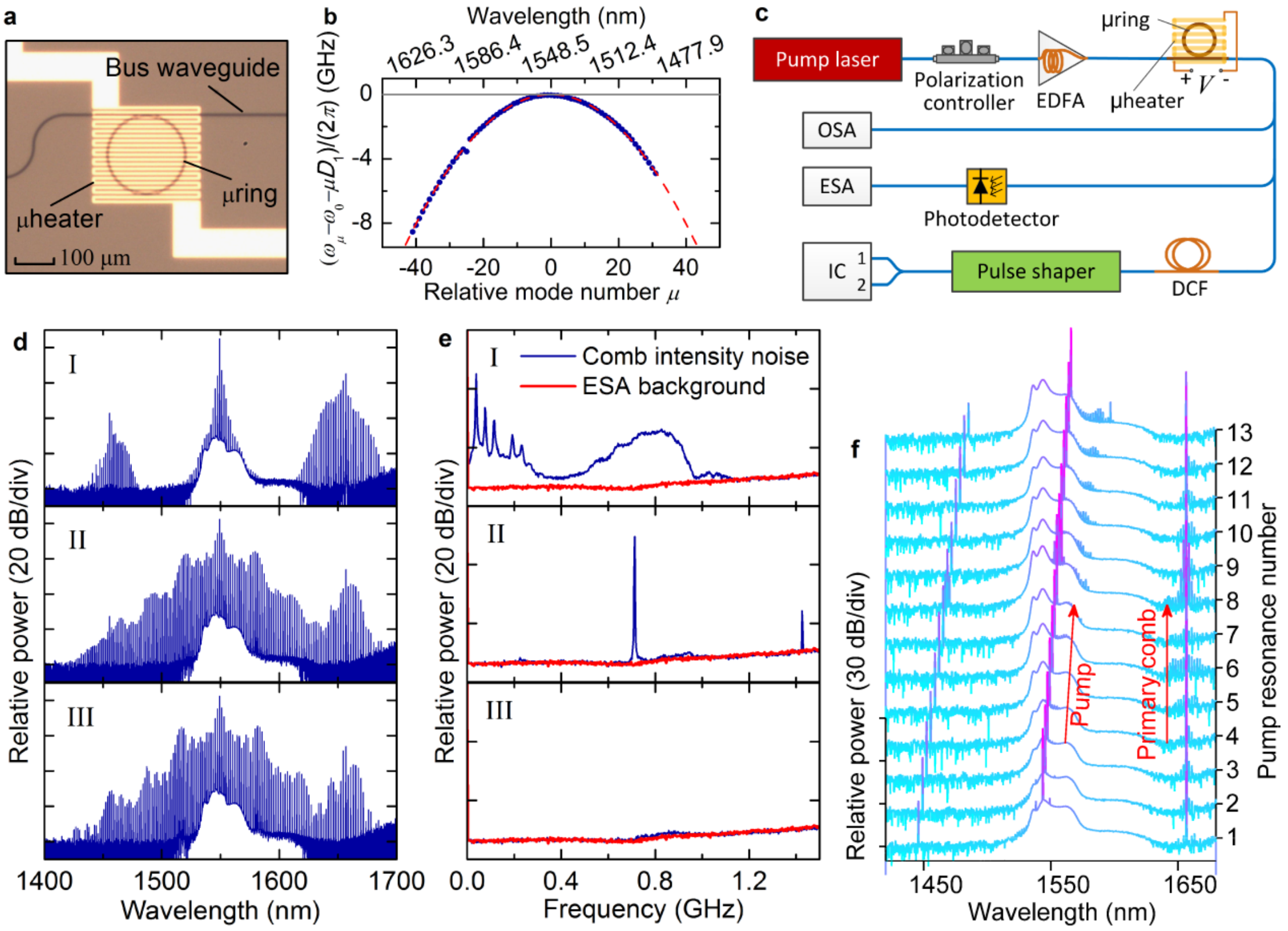

Figure 1 | Comb generation with normal-dispersion SiN microring. **a,** Microscope image of the microring. **b,** Deviation of the resonance frequencies (blue dots) $\omega_\mu = \omega_0 + D_1\mu + \frac{1}{2}D_2\mu^2 + \ldots$ from an equidistant frequency grid defined by $\omega_0 + D_1\mu$ (gray line), where $\omega_0$ is the resonance pumped and $\mu$ the relative mode number. $D_1$ is the FSR at $\omega_0$. The normal dispersion is described by $D_2 \approx -2\pi 10\ \mathrm{MHz}$ (red dashed line) while higher-order terms are negligible. **c,** Experimental setup. EDFA, Erbium-doped fiber amplifier; DCF, dispersion-compensating fiber; OSA, optical spectrum analyzer; ESA, Electrical spectrum analyzer; IC, intensity correlator. **d,** Comb generation showing a mode-locking transition. **e,** Comb intensity noise (the resolution bandwidth is 1 MHz). The pump power is injected further into the microring from I, II, to III by controlling the detuning between the pump laser and the resonance. **f,** Primary comb pinning which is a signature of mode interaction (see the main text for the comb generation details).

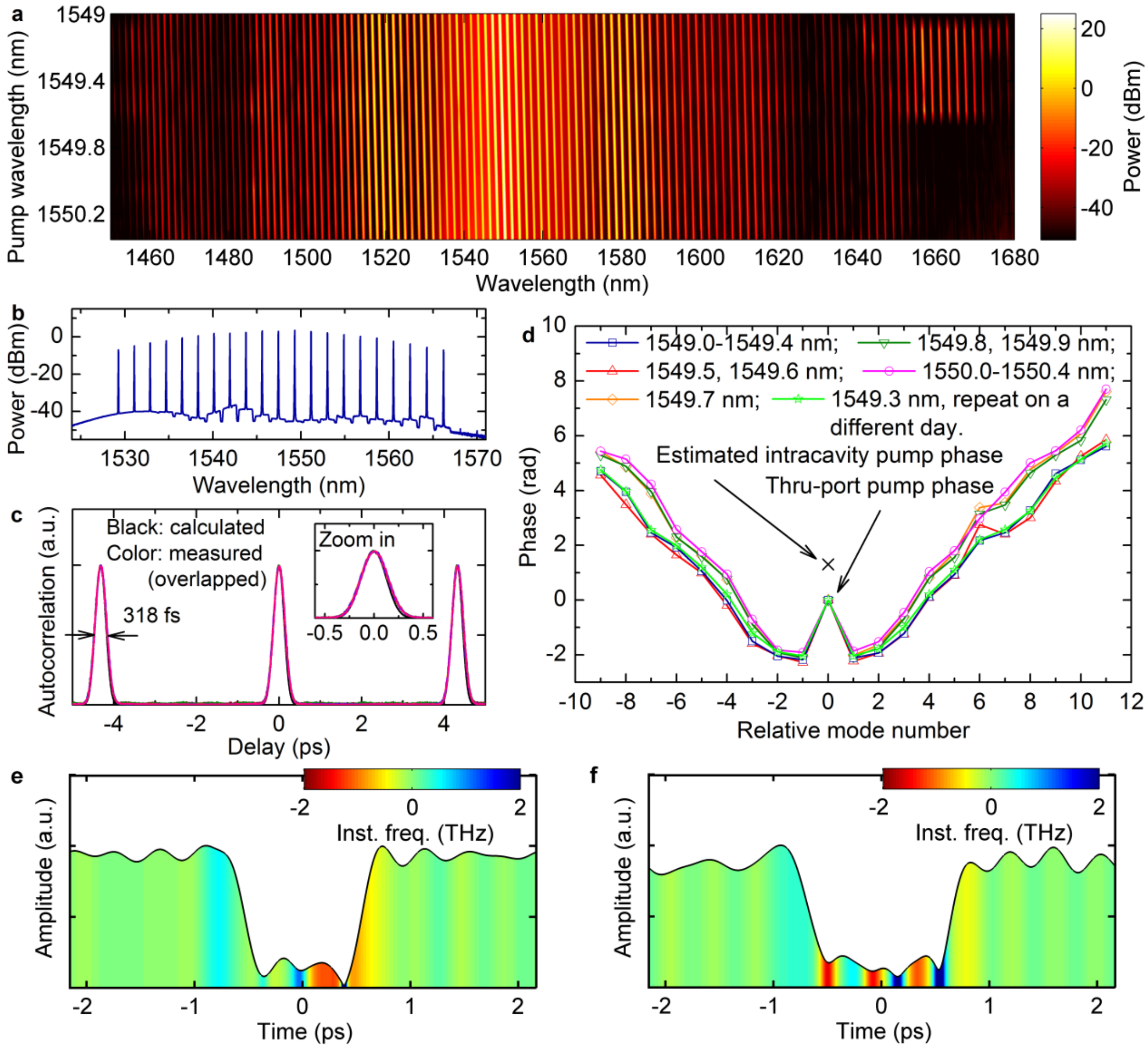

Figure 2 | Comb characterization through line-by-line shaping. **a**, Spectrum tuning of the broadband comb. The comb was initially generated with the pump wavelength 1549.3 nm, and then tuned between 1549 nm and 1550.4 nm with a step of 0.1 nm. **b**, Shaped Gaussian spectrum before autocorrelator (pumping at 1549.3 nm). **c**, Measured intensity autocorrelation when the comb phases were compensated to form a transform-limited pulse train. There are 16 curves overlapped. One is the calculated result by assuming perfect compensation; each of the other fifteen corresponds to one comb with different pump wavelengths. **d**, Retrieved comb phase using the $\exp(i\omega t)$ sign convention commonly used in ultrafast optics. The different curves correspond to different pump wavelengths. **e & f**, Reconstructed time-domain waveforms in the microresonator when the pump wavelength was 1549.3 nm and 1550.4 nm, respectively. The comb lines used for reconstruction contain 82% of the total power excluding the pump (91% including the pump).

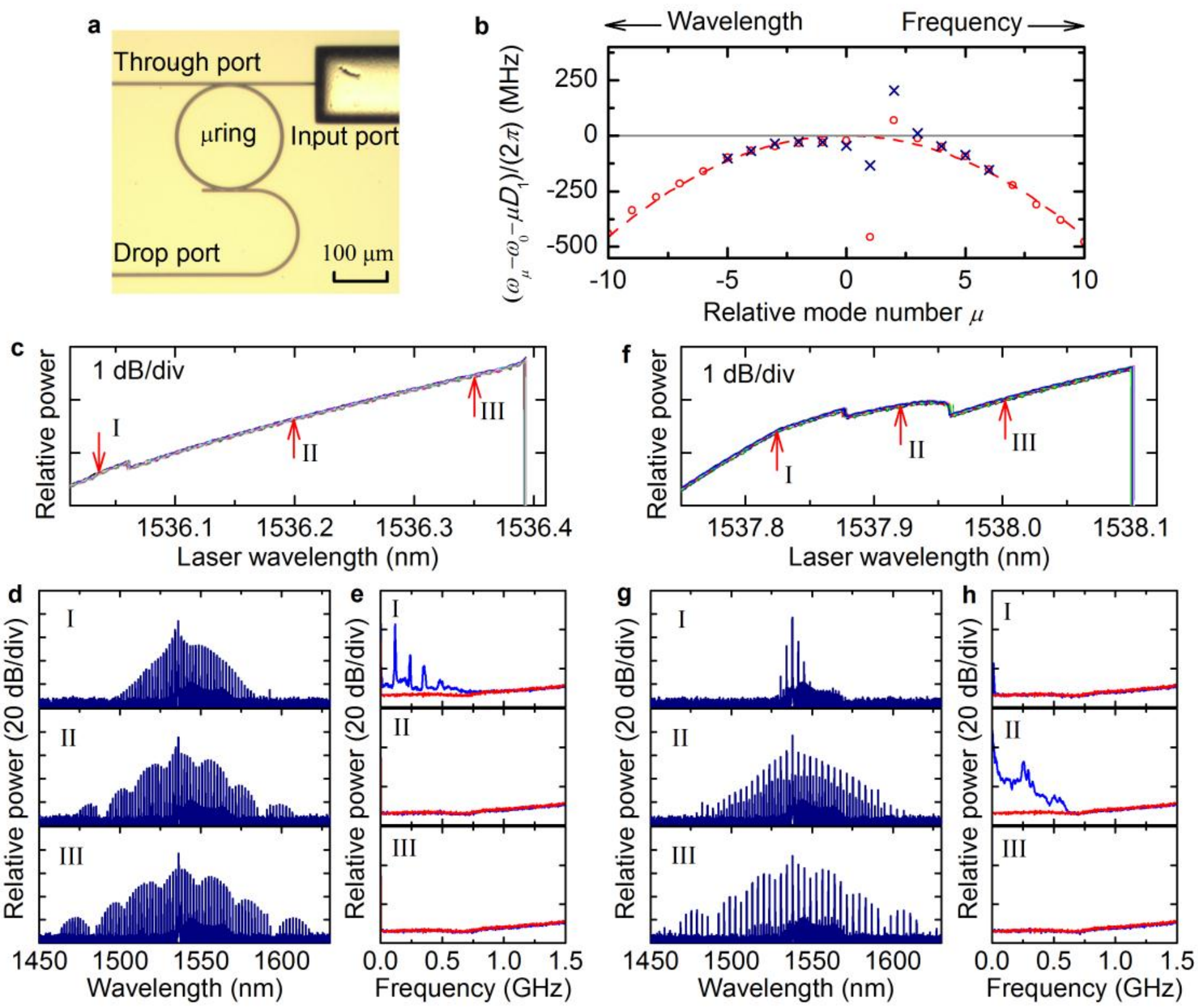

Figure 3 | Drop-port investigation of normal-dispersion combs. **a**, Microscopy image of the microring. **b**, Deviation of the resonance frequencies from an equidistant frequency grid. The definitions of symbols are similar to Fig. 1b. Mode 0 corresponds to the resonance around 1537.4 nm. Red circles: cold cavity; blue cross: pumped cavity when mode −7 was pumped with ~1 W. **c & f**, Drop-port power when mode 1 and mode 0 were pumped for comb generation respectively. Twenty measurements are closely overlapped with different colors. **d & e**, Comb spectrum and intensity noise (blue) at each stage when mode 1 was pumped. **g & h**, Comb spectrum and intensity noise (blue) at each stage when mode 0 was pumped. The background noise of the electrical spectrum analyzer (red) is also shown in **e & h**.

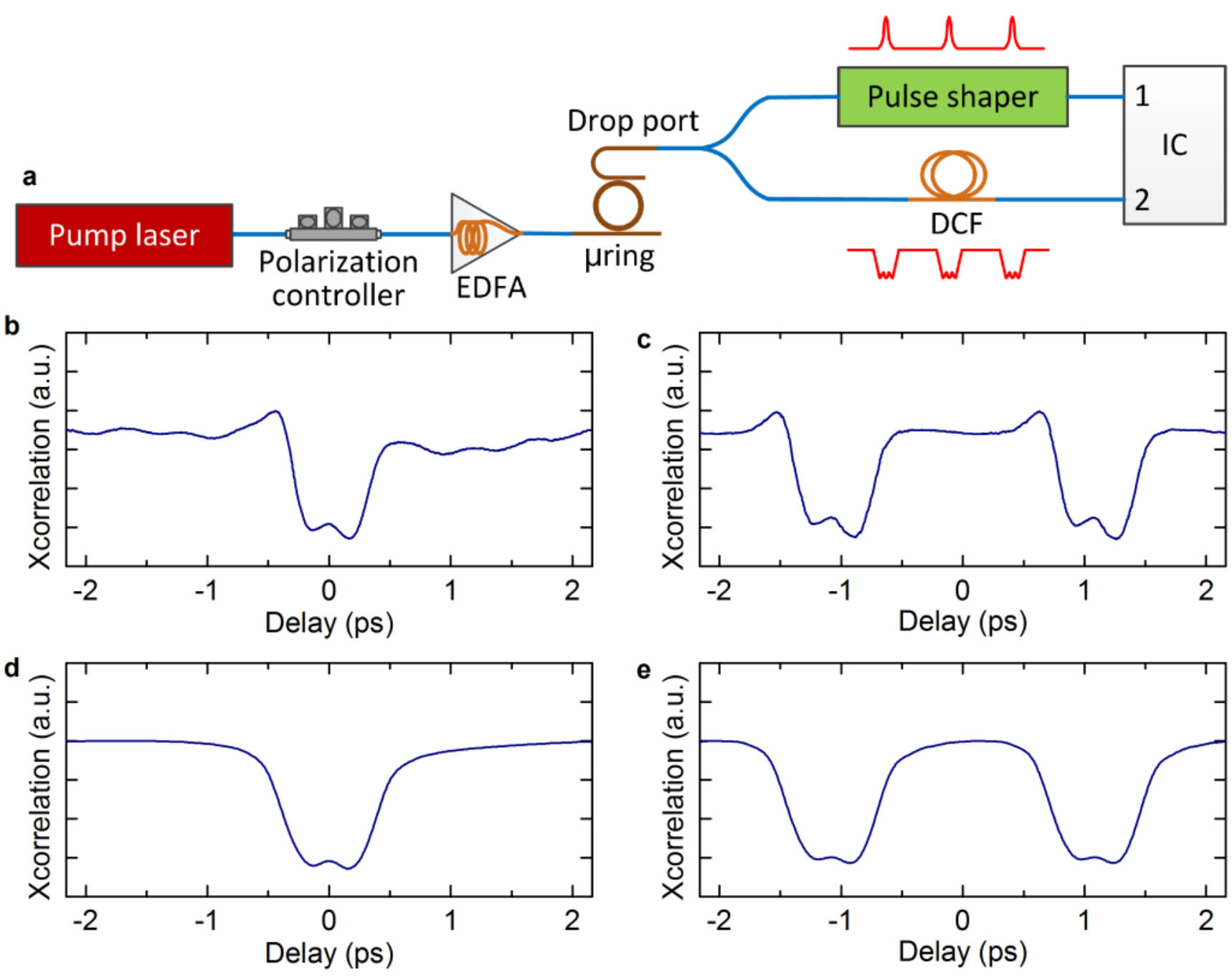

Figure 4 | Self-referenced crosscorrelation of the dark soliton combs. **a**, Experimental setup. EDFA, Erbium-doped fiber amplifier; DCF, dispersion-compensating fiber; IC, intensity correlator. **b & c**, Measured crosscorrelation for the 1-FSR comb and 2-FSR comb shown in Fig. 3d III and Fig. 3g III respectively. The width of the compressed bright pulse after pulse shaper is ~190 fs. **d & e**, Simulated crosscorrelation for the 1-FSR comb and 2-FSR comb respectively. (See Fig. 5 for the simulation details.)

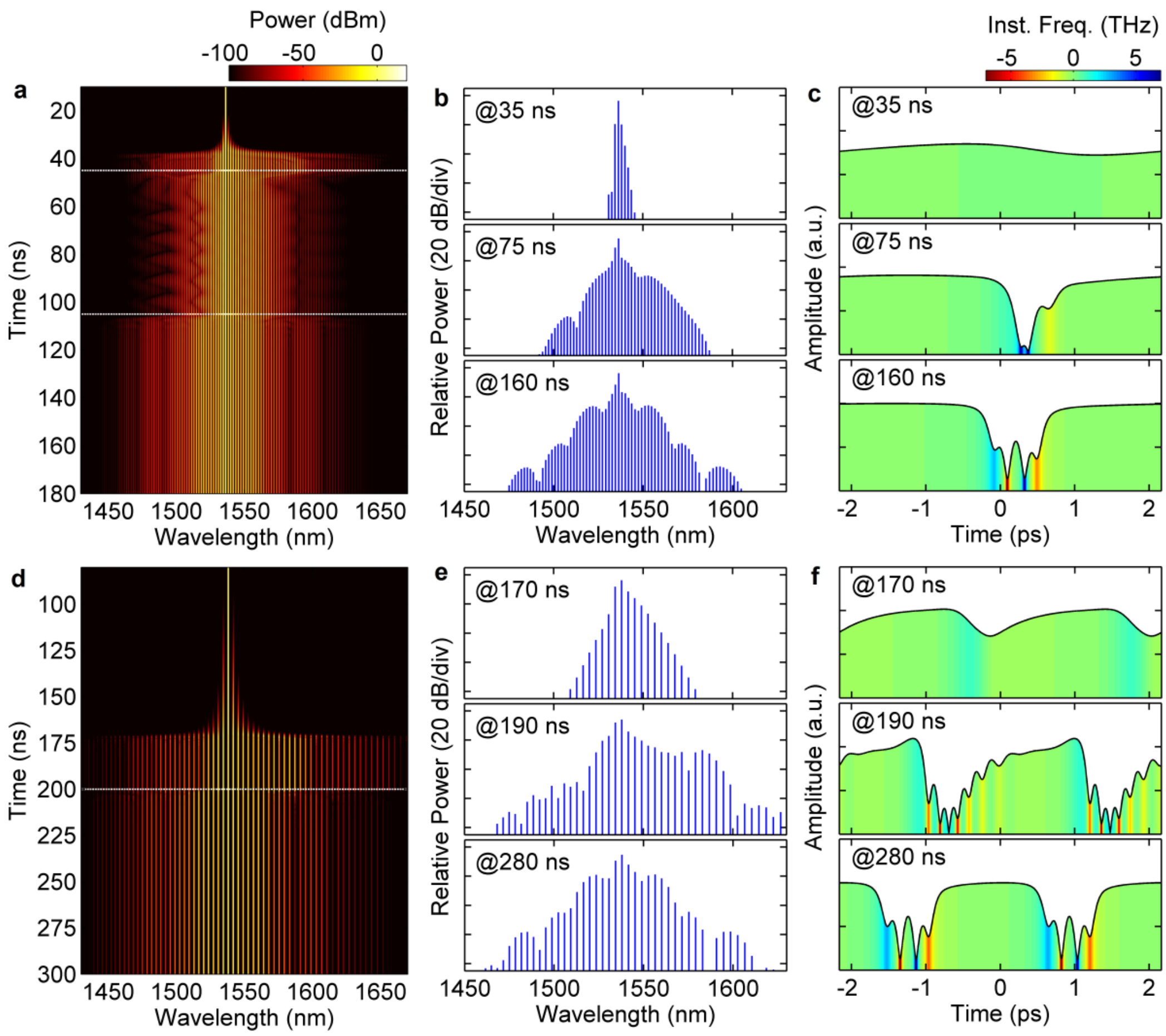

Figure 5 | Simulation of dark soliton excitation for the microring in Fig. 3a. **a,** Evolution of the comb spectrum versus the slow time when mode 1 is pumped. The initial intracavity field is the steady-state upper-branch continuous-wave solution plus weak noise. The mode interaction is taken into account by applying an additional phase shift to mode 1 and mode 2. The phase detuning is increased at slow time 45 ns and 105 ns (white dash lines). **b & c,** Transient comb spectrum and waveform at different time in subplot **a**. **d,** Evolution of the comb spectrum versus the slow time when mode 0 is pumped. The phase detuning is increased at 200 ns. **e & f,** Transient comb spectrum and waveform at different time in subplot **d**.

# Supplementary information to
# Mode interaction aided excitation of dark solitons in microresonators constructed of normal dispersion waveguides

Xiaoxiao Xue[1], Yi Xuan[1,2], Yang Liu[1], Pei-Hsun Wang[1], Steven Chen[1], Jian Wang[1,2], Dan E. Leaird[1], Minghao Qi[1,2], and Andrew M. Weiner[1,2*]

[1]*School of Electrical and Computer Engineering, Purdue University, 465 Northwestern Avenue, West Lafayette, Indiana 47907-2035, USA*

[2]*Birck Nanotechnology Center, Purdue University, 1205 West State Street, West Lafayette, Indiana 47907, USA*

**amw@purdue.edu*

## 1. Heterodyne beat note measurement

The mode-locking transitions shown in Fig. 1d, Fig. 3d and Fig. 3g in the main paper were also verified by measuring the beat note of a selected comb line with a narrow-linewidth reference laser. Figure S1 shows the results at different stages of the transition which coincide with the intensity noise measurements. The mode-locked state is indicated by the narrow linewidth of the beat note.

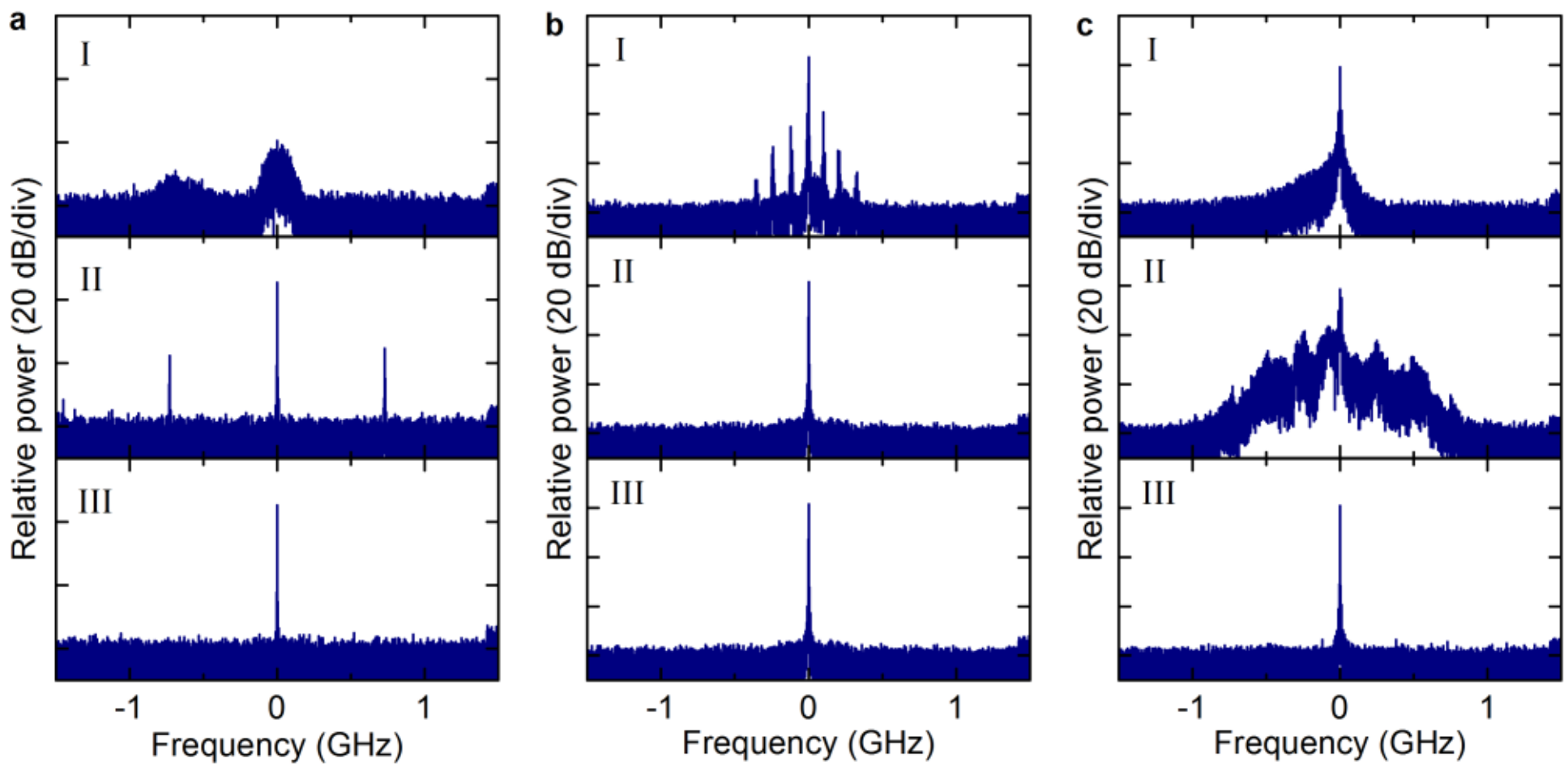

Fig. S1 Heterodyne beat note of the comb line with a narrow-linewidth reference laser. The short-term linewidth of the pump laser and the reference laser is <100 kHz and <200 kHz, respectively. The comb line tested is the first line to the red of the pump. Subplot **a** corresponds to Figs. 1d and 1e; subplot **b** corresponds to Figs. 3d and 3e; subplot c corresponds Figs. 3g and 3h. The resolution bandwidth of the electrical spectrum analyzer is 1 MHz.

## 2. Lugiato-Lefever (L-L) equation

The mean-field L-L equation is given by [S1], [S2]

$$t_R \frac{\partial E(t,\tau)}{\partial t} = \left[ -\alpha - i\delta_0 - iL\frac{\beta_2}{2}\frac{\partial^2}{\partial \tau^2} + i\gamma L \left| E(t,\tau) \right|^2 \right] E(t,\tau) + \sqrt{\theta} E_{in} \qquad \text{(S1)}$$

where $E(t,\tau)$ is the intracavity field; $t$ slow time; $\tau$ fast time; $t_R$ cavity roundtrip time;

$\alpha = (\alpha_i + \theta)/2$ roundtrip amplitude loss; $\alpha_i$ roundtrip intensity loss due to absorption and scattering in the cavity; $\theta$ coupling intensity loss (see Fig. S2); $\delta_0 = (\omega_0 - \omega_p)t_{\mathrm{R}}$ phase detuning where $\omega_p$ is the pump frequency and $\omega_0$ is the resonant frequency closest to $\omega_p$; $L$ roundtrip length; $\beta_2 = \mathrm{d}^2\beta/\mathrm{d}\omega^2|_{\omega=\omega_0}$ the second-order dispersion coefficient; $\gamma$ nonlinear Kerr coefficient; $E_{\mathrm{in}}$ external driving field (i.e., pump field).

The normalized L-L equation is given by [S1]

$$\frac{\partial F(t',\tau')}{\partial t'} = \left[-1 - i\Delta - i\eta\frac{\partial^2}{\partial \tau'^2} + i\left|F(t',\tau')\right|^2\right]F(t',\tau') + S \tag{S2}$$

The normalization is performed as follows

$$t' = \frac{\alpha t}{t_{\mathrm{R}}} \tag{S3}$$

$$\tau' = \tau\sqrt{\frac{2\alpha}{|\beta_2|L}} \tag{S4}$$

$$F = E\sqrt{\frac{\gamma L}{\alpha}} \tag{S5}$$

$$S = E_{\mathrm{in}}\sqrt{\frac{\gamma L\theta}{\alpha^3}} \tag{S6}$$

$$\Delta = \frac{\delta_0}{\alpha} \tag{S7}$$

$$\eta = \mathrm{sign}(\beta_2) \tag{S8}$$

$$\alpha' = 1 \tag{S9}$$

$$\gamma' = 1 \tag{S10}$$

where $t'$ is the slow time scaled with respect to the cavity photon lifetime; $\tau'$ normalized fast time; $F$ normalized intracavity field; $S$ normalized pump field; $\Delta$ phase detuning scaled with respect to the cavity loss; $\eta$ sign of the second-order dispersion coefficient; $\alpha'$ normalized cavity loss; $\gamma'$ normalized nonlinear Kerr coefficient.

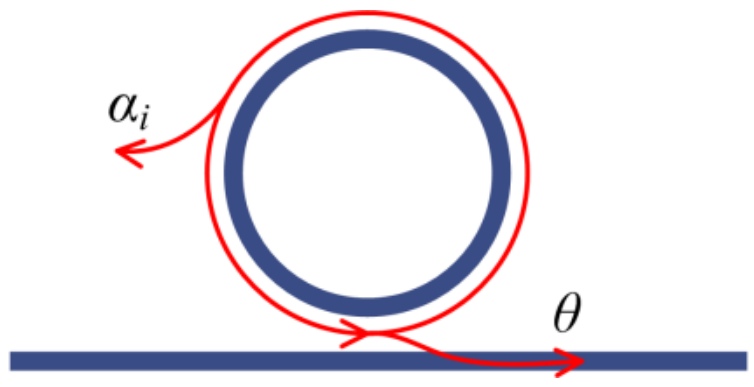

Figure S2 | Microring resonator.

### 3. Modulational instability in normal-dispersion microresonators.

The CW steady-state solutions of Eq. (S2) satisfy the well-known cubic equation [S1]

$$X = Y^3 - 2\Delta Y^2 + \left(\Delta^2 + 1\right) Y \qquad \text{(S11)}$$

where $X = |S|^2$ and $Y = |F|^2$. For $\Delta \le \sqrt{3}$, $Y(X)$ is single-valued; whereas for $\Delta > \sqrt{3}$, $Y(X)$ has the characteristic S-shape of the bistable hysteresis cycle (Fig. S3). The negative slope branch of the response curve is unstable with respect to CW perturbations, i.e., for intensities that lie in between the two values $Y_{\pm} = [2\Delta \pm (\Delta^2 - 3)^{1/2}]/3$.

It has been demonstrated that modulational instability may occur in normal-dispersion nonlinear cavities by optimizing the phase detuning and the intracavity intensity in the region [S1], [S3], [S4]

$$1 < Y < \Delta/2 \text{, with } \Delta > 2 . \qquad \text{(S12)}$$

It refers to a fraction of the lower branch near the limit point (Fig. S3). In experiments, the pump laser frequency usually starts to the blue of the cold-cavity resonance ($\Delta < 0$) and is then continuously tuned toward the red (i.e., from higher to lower frequency) to overcome resonance red-shifting caused by thermal effect and Kerr effect [S5]. This process corresponds to continuously increasing the cold-cavity phase detuning (under the sign convention in Eqs. (S1) & (S2)) between the pump frequency and the cold-state cavity while the pump power is fixed, and prevents the intracavity power getting to the modulational instability region on the lower branch (Fig. S4). Due to the Kerr effect induced resonance shift, the effective detuning may have sign different than the cold-cavity detuning. The upper branch is effectively blue detuned while the lower branch is effectively red detuned [S5]. The modulational instability region on the lower branch is generally thermally unstable for microresonators [S6]. Although it might be possible to get to the modulational instability region by controlling the phase detuning and the pump power in tandem, experiments and simulations have shown that the intracavity power may switch to the upper branch due to the instability of Turing patterns in this region [S7]. No dark soliton formation has ever been found in this process.

In the case of mode interaction, modulational instability may occur on the upper branch [S8]-[S9]. In experiments, it provides an easy way to get mode-interaction-aided initial comb lines which may act as a source for exciting dark solitons.

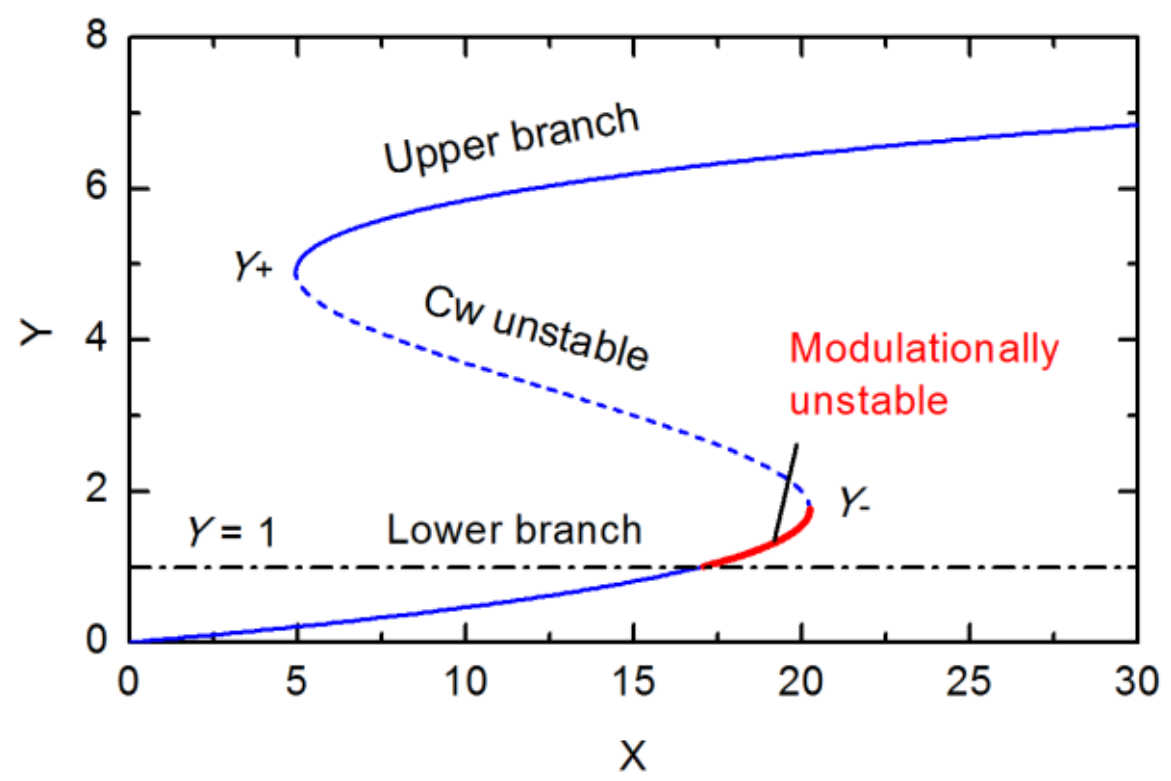

Figure S3 | Steady-state intracavity intensity versus driving intensity when the phase detuning $\Delta = 5$.

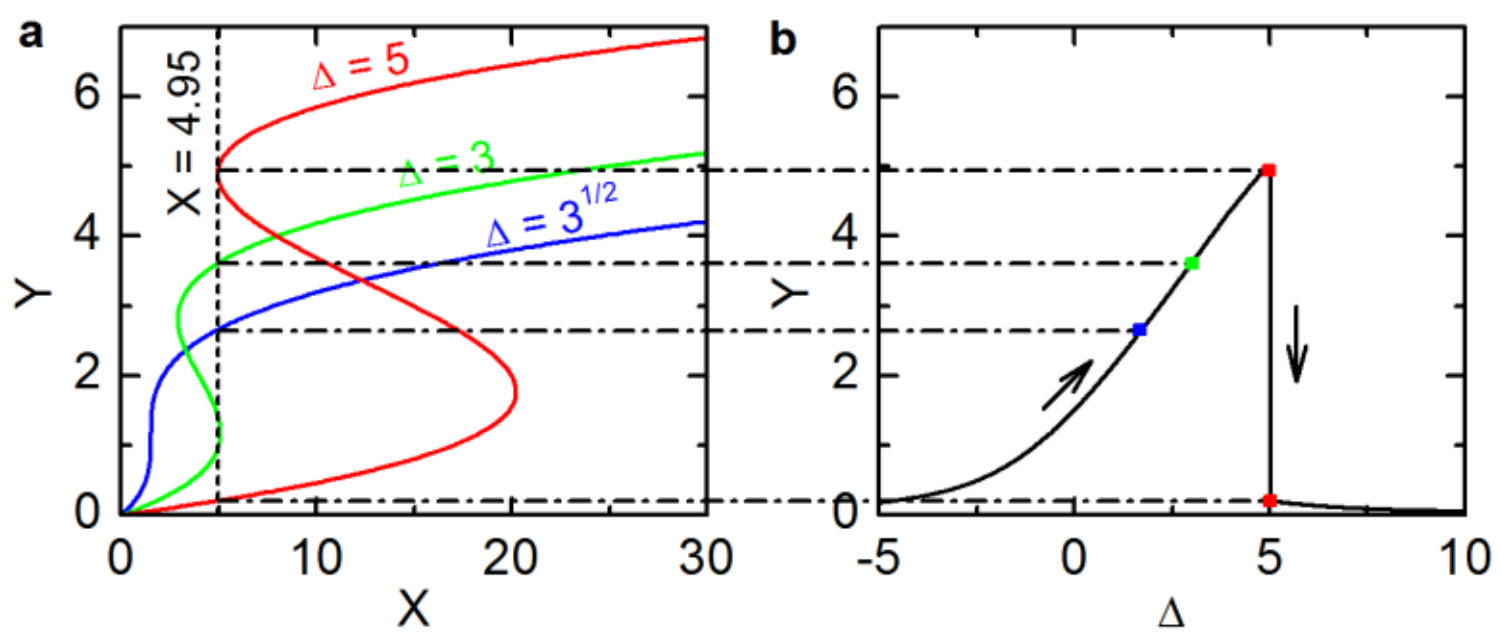

Figure S4 | Intracavity intensity when the cold-cavity phase detuning is continuously increased. **a**, Intracavity intensity versus driving intensity for $\Delta = \sqrt{3}, 3, 5$. **b**, Intracavity intensity versus phase detuning. The driving intensity is 4.95. The intracavity intensity stays on the upper branch before it drops down to the lower branch.

## 4. Pump line in the cavity

Pulse shaping experiments performed on the output comb field allow us to determine the waveform at the output waveguide (usually the through port). In order to reconstruct the comb field internal to the microresonator, we need to relate the internal pump field to the externally measured field. This takes some effort, as the pump line at the through port is the coherent summation of the output coupled fraction of the internal pump field with the input field transmitted directly to the through port. In the following we first outline a procedure whereby the phase of the internal pump field is obtained from the amplitudes of the internal comb lines; we test the obtained relation by comparing with simulation results (section 4.1). Then in section 4.2, we show how to use the comb spectrum measured at the through port, together with the results of section 4.1, to determine the internal pump field. Experimental results validating the procedure are shown in section 4.3. Sections 4.4 and 4.5 described methods for extracting the linear cavity parameters and identifying the effective detuning region, respectively.

### 4.1. Retrieving pump phase from the comb spectral amplitude

From the perspective of the pump line, comb generation will introduce an additional loss to the pump line. According to energy conservation, the effective cavity loss for the pump line is

given by

$$\alpha_{\text{eff}} = \frac{\alpha \cdot P_{\text{all comb lines}}^{\text{cavity}}}{P_{\text{pump}}^{\text{cavity}}} \tag{S13}$$

where $P_{\text{all comb lines}}^{\text{cavity}}$ is power of all the comb lines in the cavity; $P_{\text{pump}}^{\text{cavity}}$ is power of the pump line. Suppose that the effective phase detuning for the pump line is denoted as $\delta_{\text{eff}}$, the intracavity pump line is then expressed as

$$E_{\text{pump}}^{\text{cavity}} = \frac{\sqrt{\theta} E_{\text{in}}}{\alpha_{\text{eff}} + i\delta_{\text{eff}}} \tag{S14}$$

It is worth noting that $\delta_{\text{eff}}$ is different from $\delta_0$ in Eq. (S1) which represents the phase detuning between the pump frequency and the cold-state resonance. Here $\delta_{\text{eff}}$ represents the phase detuning between the pump frequency and the shifted resonance. (The resonance is shifted due to Kerr effect.) Simulations (see for example Fig. S5) show that for bright soliton formation, the pump frequency is red detuned with respective to the shifted resonance (i.e., $\delta_{\text{eff}} > 0$, as reported in [S5]); whereas for dark soliton formation, the pump frequency is blue detuned (i.e., $\delta_{\text{eff}} < 0$).

Based on the amplitude information of the comb spectrum and the coupling condition of the cold-state cavity, one can calculate the phase of the pump line without knowing the phase of the other comb lines or the exact time-domain waveform. Once the stable comb spectral amplitude as well as the detuning region is known, $\delta_{\text{eff}}$ can be obtained by solving the following equation

$$\left| E_{\text{pump}}^{\text{cavity}} \right| = \left| \frac{\sqrt{\theta} E_{\text{in}}}{\alpha_{\text{eff}} + i\delta_{\text{eff}}} \right| \tag{S15}$$

The phase of the pump line can then be calculated by substituting $\delta_{\text{eff}}$ into Eq. (S14).

Numerical simulations are performed based on the normalized L-L equation (Eq. (S2)). In normalized form, equations (S13) and (S14) are given by

$$\alpha_{\text{eff}}^{'} = \frac{P_{\text{all comb lines}}^{\text{cavity}}}{P_{\text{pump}}^{\text{cavity}}} \tag{S16}$$

$$F_{\text{pump}}^{\text{cavity}} = \frac{S}{\alpha_{\text{eff}}^{'} + i\Delta_{\text{eff}}} \tag{S17}$$

Figure S5 shows the simulation results. The pump phase retrieved according to Eqs. (S13)-(S15) agrees well with the actual pump phase.

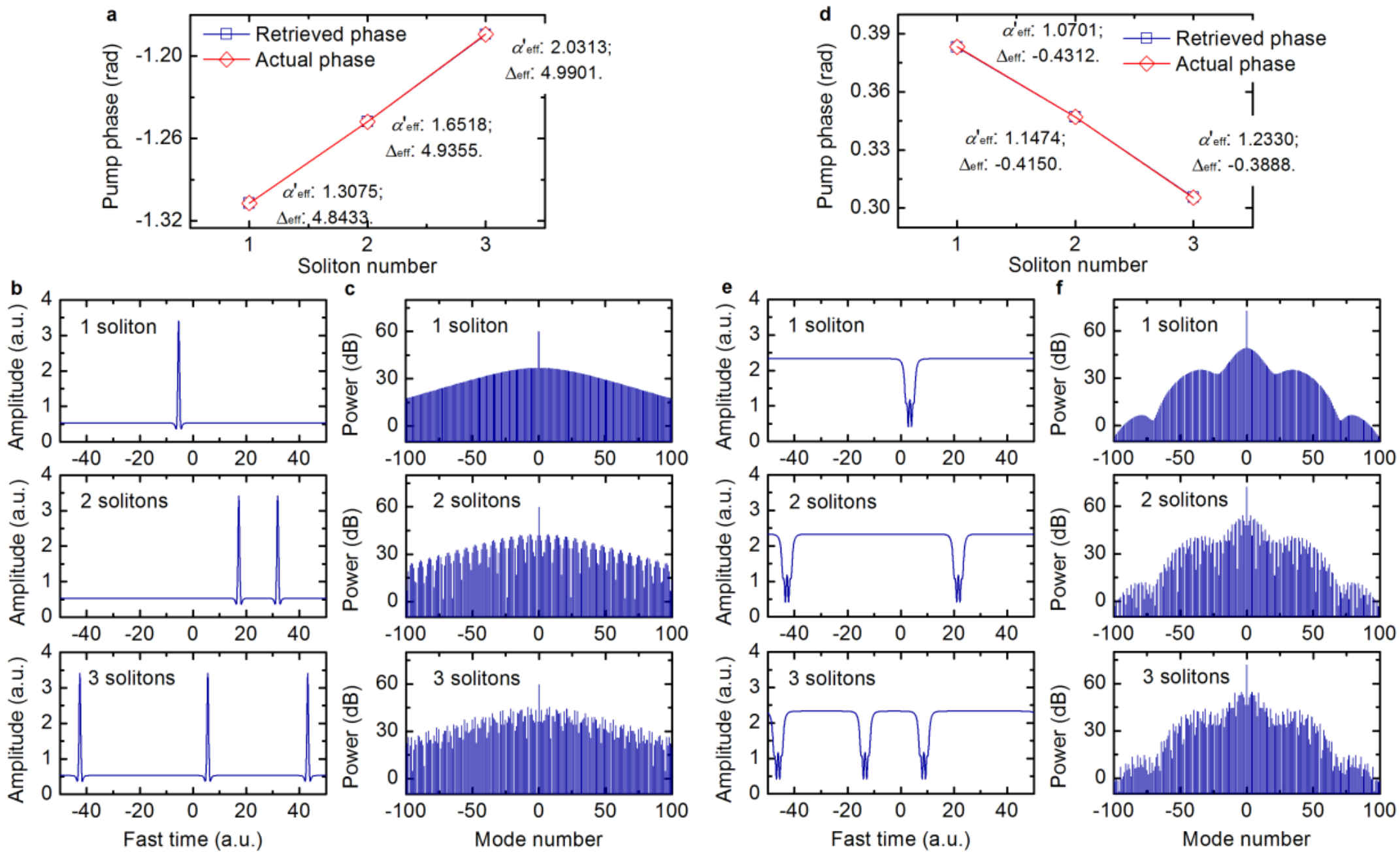

Figure S5 | Numerical simulations of retrieving the pump phase from the comb spectral amplitude. **a-c**, Bright solitons. Three cases corresponding to 1, 2, 3 solitons are shown. The time-domain waveforms and frequency-domain comb spectra are shown in (**b**) and (**c**) respectively. **d-f**, Dark solitons. Three cases corresponding to 1, 2, 3 solitons are shown. The time-domain waveforms and frequency-domain comb spectra are shown in (**e**) and (**f**) respectively. The simulation parameters are as follows: $S = 2.55$, $FSR = 0.01$, $\Delta = 5$, $\eta = -1$ for subplots **a**-**c**, $\eta = 1$ for subplots **d**-**f**. The distance between the initial bright or dark pulses is randomly selected.

## 4.2. Correction of the pump line measured at through port

When the comb spectrum is measured at the through port, the pump line consists of two components: one from coupling out of the cavity, the other from the bus waveguide. To reconstruct the time-domain waveform, the measured pump line at the through port should be corrected to represent the component from the cavity.

Suppose that the correction factor is denoted as

$$C = \frac{E_{\text{pump}}^{\text{from cavity}}}{E_{\text{pump}}^{\text{total}}} = \frac{E_{\text{pump}}^{\text{from cavity}}}{E_{\text{pump}}^{\text{from cavity}} + E_{\text{pump}}^{\text{from bus}}} \tag{S18}$$

where $E_{\text{pump}}^{\text{from cavity}}$ and $E_{\text{pump}}^{\text{from bus}}$ represent the component from the cavity and the bus waveguide respectively. The effective cavity loss for the pump line is then

$$\alpha_{\text{eff}} = \frac{\alpha \cdot \left( |C|^2 P_{\text{pump}}^{\text{total}} + P_{\text{other comb lines}} \right)}{|C|^2 P_{\text{pump}}^{\text{total}}} \tag{S19}$$

where $P_{\text{pump}}^{\text{total}} = \left|E_{\text{pump}}^{\text{total}}\right|^2$ pump power measured at the through port; $P_{\text{other comb lines}}$ power of the other comb lines.

The pump component from coupling out of the cavity is given by

$$E_{\text{pump}}^{\text{from cavity}} = -\frac{\theta E_{\text{in}}}{\alpha_{\text{eff}} + i\delta_{\text{eff}}} \tag{S20}$$

And the pump component from the bus waveguide

$$E_{\text{pump}}^{\text{from bus}} = \sqrt{1-\theta}E_{\text{in}} \tag{S21}$$

The amplitude drop of the pump line in comb generation compared to when the pump frequency is out of resonance can be measured in experiments and is given by

$$|L| = \left|\frac{E_{\text{pump}}^{\text{from cavity}} + E_{\text{pump}}^{\text{from bus}}}{E_{\text{pump}}^{\text{from bus}}}\right| = \left|\frac{-\frac{\theta}{\alpha_{\text{eff}} + i\delta_{\text{eff}}} + \sqrt{1-\theta}}{\sqrt{1-\theta}}\right| \tag{S22}$$

The correction factor is given by

$$C = \frac{-\frac{\theta}{\alpha_{\text{eff}} + i\delta_{\text{eff}}}}{-\frac{\theta}{\alpha_{\text{eff}} + i\delta_{\text{eff}}} + \sqrt{1-\theta}} \tag{S23}$$

Equations (S22) & (S23) are two independent equations for $\delta_{\text{eff}}$ and $|C|$ and can be numerically solved. For the microring shown in Fig. 1a of the main paper, the parameters are as follows

$\alpha = 3.3978\times10^{-3}$;
$\theta = 5.3613\times10^{-3}$;
$P_{\text{other comb lines}} / P_{\text{pump}}^{\text{total}} = 0.8937$;

$|L| = 0.5957$;

The retrieved effective phase detuning, effective cavity loss, and correction factor are

$\delta_{\text{eff}} = -3.2281\times10^{-3}$;

$\alpha_{\text{eff}} = 4.5623\times10^{-3}$;

$C = 0.3946 - 1.5659i$ (corresponding to amplitude 4.2 dB and phase −1.3 rad).

In reconstructing the time-domain waveform in the cavity, the pump amplitude measured at the through port is increased by 4.2 dB and the phase is shifted by 1.3 rad (which is inverse of the phase of $C$ because of the different sign conventions we use in simulations and experiments).

### 4.3. Experimental results

Here we show experimental results which validate the through-port pump correction method outlined above. The same microring shown in Fig. 3a of the main paper was pumped with ~0.8 W. Since the pumped power is reduced compared to that in Fig. 3 of the main paper, the comb has a narrower spectrum and different features. The comb spectrum also becomes more asymmetric with more comb power shifted to the longer wavelength range which falls in the passband of our pulse shaper. The comb was characterized both at the through port and the drop port. With the through port pump correcting method, the estimated complex pump amplitude gets very close to the drop port value. For the 1-FSR comb, the intensity agreement is improved from 8.14 dB (without correction procedure) to 0.73 dB; the phase agreement is improved from –2.19 rad to 0.01 rad. For the 2-FSR comb, the intensity agreement is improved from 4.25 dB to 0.58 dB; the phase agreement is improved from –2.26 rad to 0.14 rad. The reconstructed intracavity time-domain waveforms from the through-port (Figs. S6 b & f) and the drop port (Figs. S6 d & h) are very close to each other. The agreement validates our procedure to correct for the strong superimposed pump field in through port measurements (e.g., Fig. 2d of the main paper).

The parameters used in pump correction are as follows.

$\alpha = 3.0981\times10^{-3}$ ,

$\theta = 1.9327\times10^{-3}$ .

For the 1-FSR comb,

$P_{\text{other comb lines}} \big/ P_{\text{pump}}^{\text{total}} = 0.0879$ ;

$|L| = 0.7713$ ;

$\delta_{\text{eff}} = -3.6784\times10^{-3}$ ;

$\alpha_{\text{eff}} = 4.5981\times10^{-3}$ ;

$C = -0.2498 - 0.3450i$ .

For the 2-FSR comb,

$P_{\text{other comb lines}} \big/ P_{\text{pump}}^{\text{total}} = 0.1439$ ;

$|L| = 0.6446$ ;

$\delta_{\text{eff}} = -1.9669\times10^{-3}$ ;

$\alpha_{\text{eff}} = 4.1363\times10^{-3}$ ;

$C = -0.4887 - 0.4365i$ .

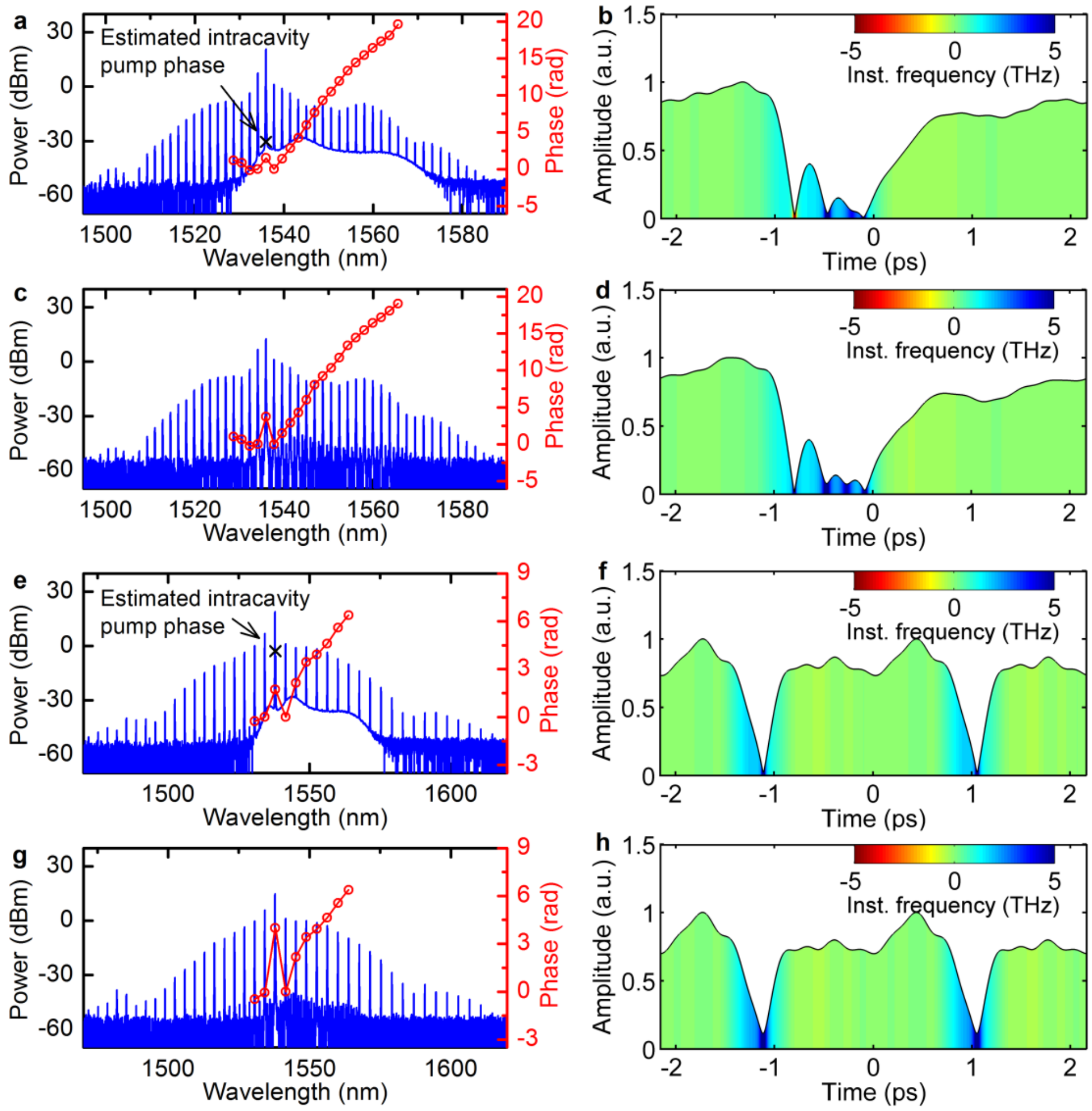

Figure S6 | Experimental results of correcting the through-port complex pump. **a – d,** 1-FSR comb**. e – h,** 2-FSR comb. **a & e,** Comb spectrum measured at through port. **c & g,** Comb spectrum measured at drop port. **b & f,** Reconstructed time-domain waveform based on through-port data. **d & h,** Reconstructed time-domain waveform based on drop-port data.

### 4.4. Extracting cavity parameters

The cavity loss $\alpha$ and the bus waveguide coupling coefficient $\theta$ can be extracted from the measured loaded Q and through-port extinction ratio (EXR), i.e. solving the following equations

$$Q_l \approx \frac{\omega_0 t_R}{2\alpha}, \tag{S24}$$

$$\mathrm{EXR} \approx \frac{\alpha}{\alpha - \theta}, \tag{S25}$$

where $\omega_0$ is the resonance frequency and $t_R$ is the round-trip time. The EXR is defined as the complex amplitude when the pump is out of resonance over that when the pump is in the resonance center. EXR is positive (negative) when the cavity is under (over) coupled.

The coupling condition can be characterized by measuring the phase response of the microresonator. Figure S7 shows the experimental setup of sideband sweeping method. The tunable laser is tuned close to the resonant frequency, and is modulated by a radiofrequency signal through single-sideband modulation. By sweeping the radiofrequency, the sideband sweeps across the resonance of the microresonator. The response of the microresonator is then transferred to the electrical domain through beating of the sideband with the carrier. Figure S7b shows the measured response of the microring shown in Fig. 1a of the main paper. The phase curve indicates that the resonator is over-coupled. Figure S7c shows the measured result of the microring used for comb generation in Fig. 3a of the main paper. Since this microring has a drop port which introduces an additional coupling loss comparable to the through port, the phase curve indicates an under-coupled condition.

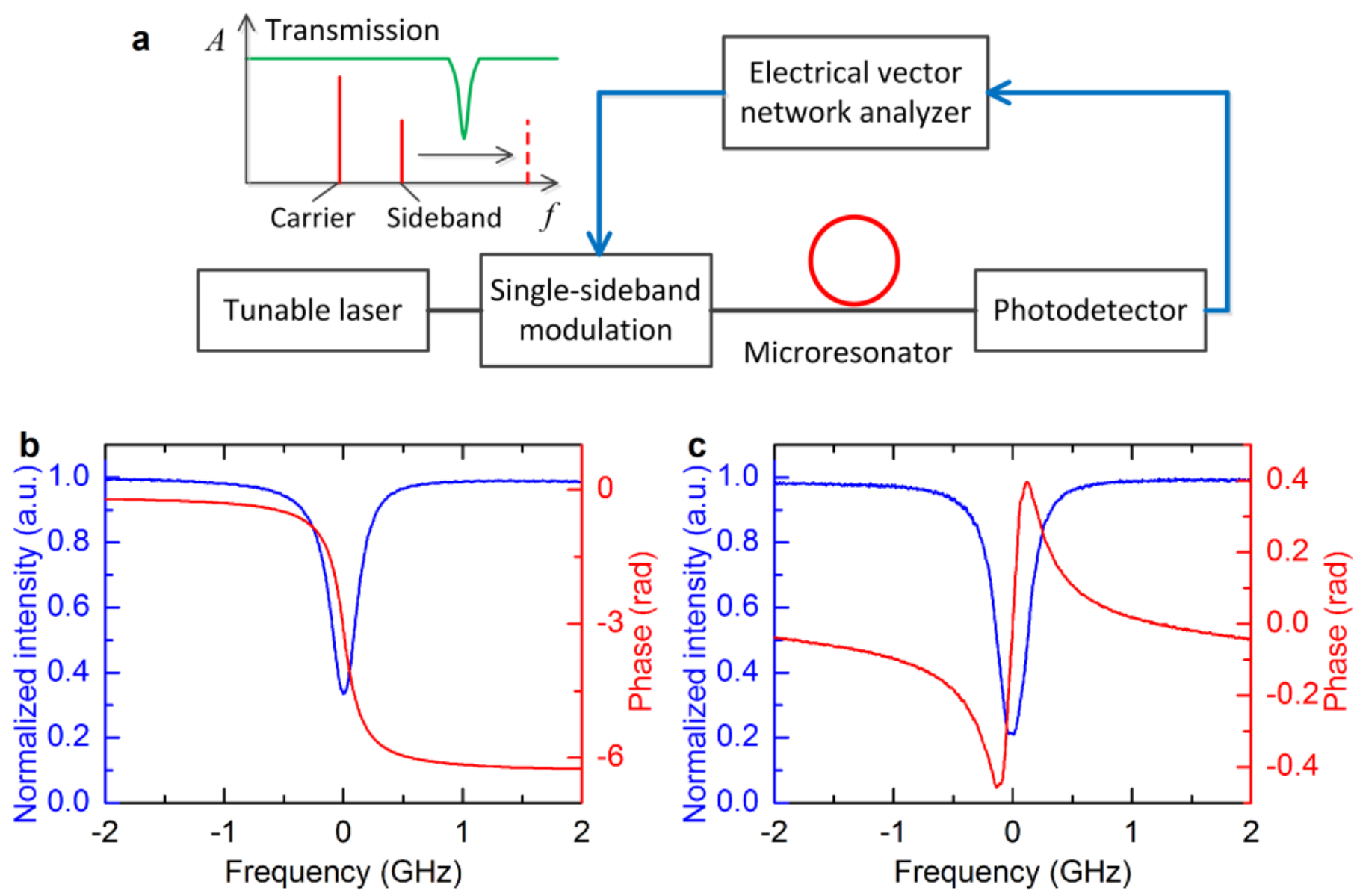

Figure S7 | **a**, Experimental setup for measuring the phase response of the microresonator. **b & c** show the results for the microring in Fig. 1a and the microring in Fig. 3a of the main paper, respectively.

### 4.5. Identifying the effective detuning region

In Fig. S6, the estimated intracavity pump phase based on the blue detuning condition agrees well with the actual value measured at drop port. The good agreement confirms that the effective detuning is indeed in the blue region. If effective red detuning is assumed, poor agreement is obtained. For the microring with no drop port shown in Fig. 1a of the main paper, the effective detuning can be identified by using the method shown in Fig. S8a which is in principle similar to the method of detecting the Pound–Drever–Hall (PDH) signal employed in [S5]. A dithering voltage is applied on the microheater. The pump line power after the microring is modulated because of the slight shift of the resonance. The detuning region can be distinguished by comparing the phases of the dithering signal ($\tilde{V}_d$) and the pump line variation (converted to a voltage $\tilde{V}_o$ through a photodetector). Example curves of $\tilde{V}_d$ and $\tilde{V}_o$ for the cold cavity (microring shown in Fig. 1a of the main paper) are shown in Fig. S8b. When the pump laser wavelength is shorter than the resonant wavelength (i.e., blue detuned), $\tilde{V}_o$ is in phase with $\tilde{V}_d$. In the other case of red detuned region, $\tilde{V}_o$ is out of phase with $\tilde{V}_d$. The measured result for the pumped cavity under dark soliton action (example comb spectrum shown in Fig. 1d III of the main paper) is shown in Fig. S8c. The curve of $\tilde{V}_o$ is in phase with $\tilde{V}_d$ suggesting an effectively blue detuned region.

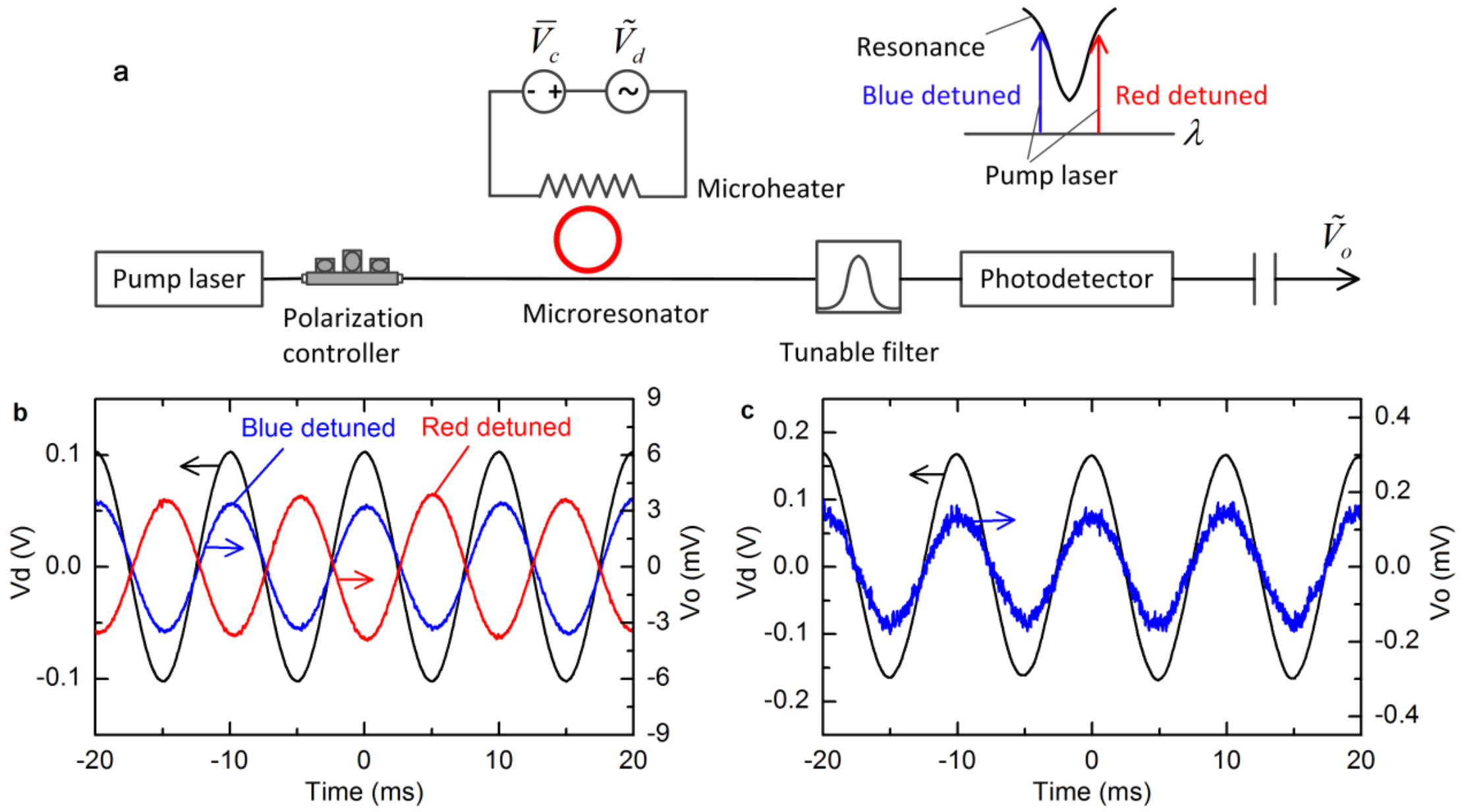

Figure S8 | Identifying the detuning region. **a,** Experimental setup. The tunable filter after microresonator is used to select the pump line in comb generation. By changing the dc voltage ($\bar{V}_c$) applied on the microheater, the detuning can be changed to either blue detuned (pump laser wavelength shorter than resonant wavelength) or red detuned (pump laser wavelength longer than resonant wavelength). **b,** Example curves of $\tilde{V}_d$ and $\tilde{V}_o$ for the cold cavity (microring shown in Fig. 1a of the main paper) measured under low pump power, showing that $\tilde{V}_o$ is in phase (out of phase) with $\tilde{V}_d$ in the blue (red) detuned region. The frequency of the dithering voltage ($\tilde{V}_d$) is 100 Hz. **c,** Measured result for the pumped cavity in comb generation. $\tilde{V}_o$ is in phase with $\tilde{V}_d$ suggesting an effectively blue detuned region. The pump wavelength is 1549.3 nm. The measurement is done after the comb transitions to a low-noise mode-locked state related to dark soliton formation (example comb spectrum shown in Fig. 1d III of the main paper).

## 5. Complex structure of dark solitons

Figure S9 shows the retrieved comb phase and reconstructed time-domain waveform measured over a larger spectral range for the dark soliton in Fig. 2 of the main paper. Here a Finisar WaveShaper 4000S which can operate in both the lightwave C and L bands was used to shape the comb, allowing access to more comb lines. Compared to Fig. 2e, here the extra comb bandwidth used in the reconstruction yields sharper edges and stronger frequency modulation. The chirped ripples at the bottom of the dark soliton also become more pronounced. Unlike bright solitons in the anomalous dispersion region, dark solitons can have complex and quite distinct features.

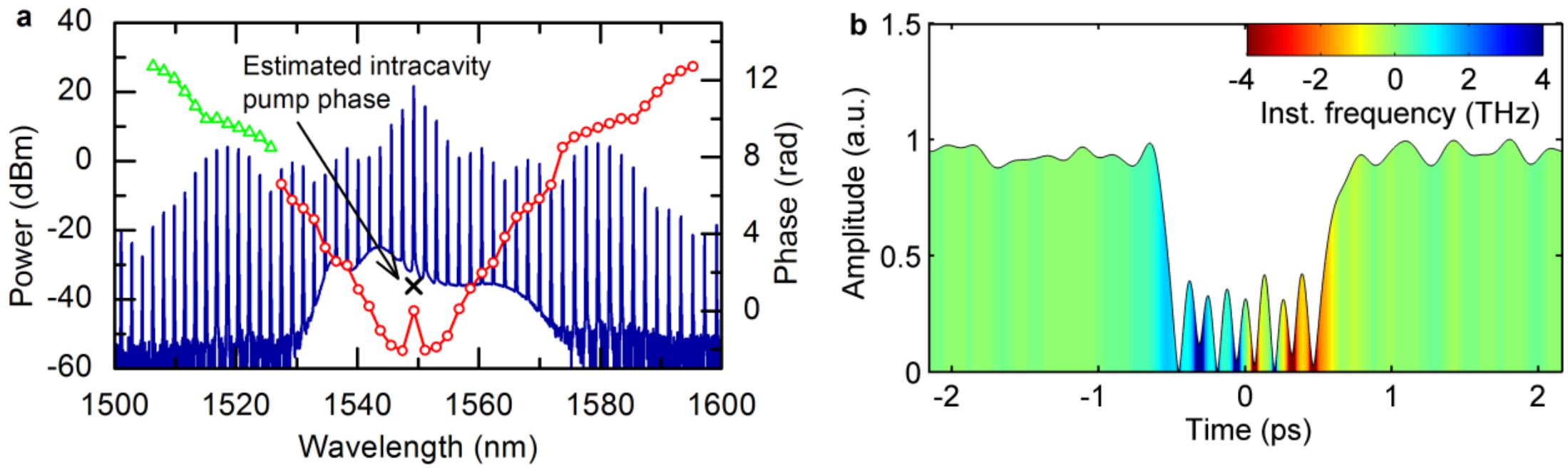

Figure S9 | Characterization of the dark soliton shown in Fig. 2 with a larger pulse shaping range. The pump wavelength is 1549.3 nm. **a,** Comb spectrum and phase. The red circles are the comb phases retrieved experimentally. The green triangles correspond to additional comb lines that fall outside of the pulse shaper operating band, for which we assume phases based on symmetry about the pump line. **b,** Reconstructed time-domain waveform. The comb lines used for reconstruction contain 99.9% of the total comb power excluding the pump (comb lines with green triangle phases based on a symmetry assumption account for 7% of the power excluding pump).

Numerical simulations are performed to model the comb generation behavior for this microring. The pump power is 0.3 W. The initial intracavity field is the steady-state continuous-wave solution on the upper branch of the bistability curve plus weak noise (~ 1 pW/mode). The phase detuning at the beginning is $3\times10^{-2}$ rad; and an additional phase shift of -0.815 rad per roundtrip is applied to modes -54 and -55. Figure S10a shows the evolution of the comb spectrum versus the slow time; figures S10b and S10c show the transient comb spectrum and time-domain waveform at different time. The comb grows up and shows some random variations with the slow time, corresponding to a high intensity noise which is similar to the experimental observation (see Fig. 1e I of the main paper). The phase detuning is increased to $3.75\times10^{-2}$ rad after 60 ns; and the additional phase shift applied to modes -54 and -55 is set to 0. The field then evolves to a stable dark pulse. The width of the dark pulse is ~1.1 ps which agrees well with the experimental result.

Please note that the L-L equation generally uses an $\exp(-i\omega t)$ convention [S1]-[S2], a convention which we adopt also in our simulations. However, for our experimental measurements we use the $\exp(i\omega t)$ convention prevalent in ultrafast optics [S14]. Therefore, in comparing simulation with experiment, one must take into account the opposite sign conventions. For our paper we choose to plot all the spectral phases using the $\exp(i\omega t)$ convention. In particular:

- In the ultrafast optics convention, we have

$$e(t)=\mathrm{Re}\left(a(t)e^{i\omega_o t}\right)=\frac{1}{2}\left[a(t)e^{i\omega_o t}+a^*(t)e^{-i\omega_o t}\right]=\frac{1}{2}\left[|a(t)|e^{i(\omega_o t+\phi(t))}+|a(t)|e^{-i(\omega_o t+\phi(t))}\right]$$

- In the convention used for L-L equation, we have

$$e(t)=\mathrm{Re}\left(a(t)e^{-i\omega_o t}\right)=\frac{1}{2}\left[a(t)e^{-i\omega_o t}+a^*(t)e^{i\omega_o t}\right]=\frac{1}{2}\left[|a(t)|e^{-i(\omega_o t-\phi(t))}+|a(t)|e^{i(\omega_o t-\phi(t))}\right]$$

In both expressions above we are using $a(t)$ as the complex envelope function, with $t$ representing fast time, a terminology common in ultrafast optics [S14]. So if we obtain $\phi(t)$ from the L-L equation, we need to change it to $-\phi(t)$ if we wish to compare it to experimental data which we plot using the prevalent ultrafast optics convention. Equivalently we can say if we have $a(t)$ from the L-L equation, we need to change it to $a^*(t)$ if we want to be consistent with the ultrafast optics convention.

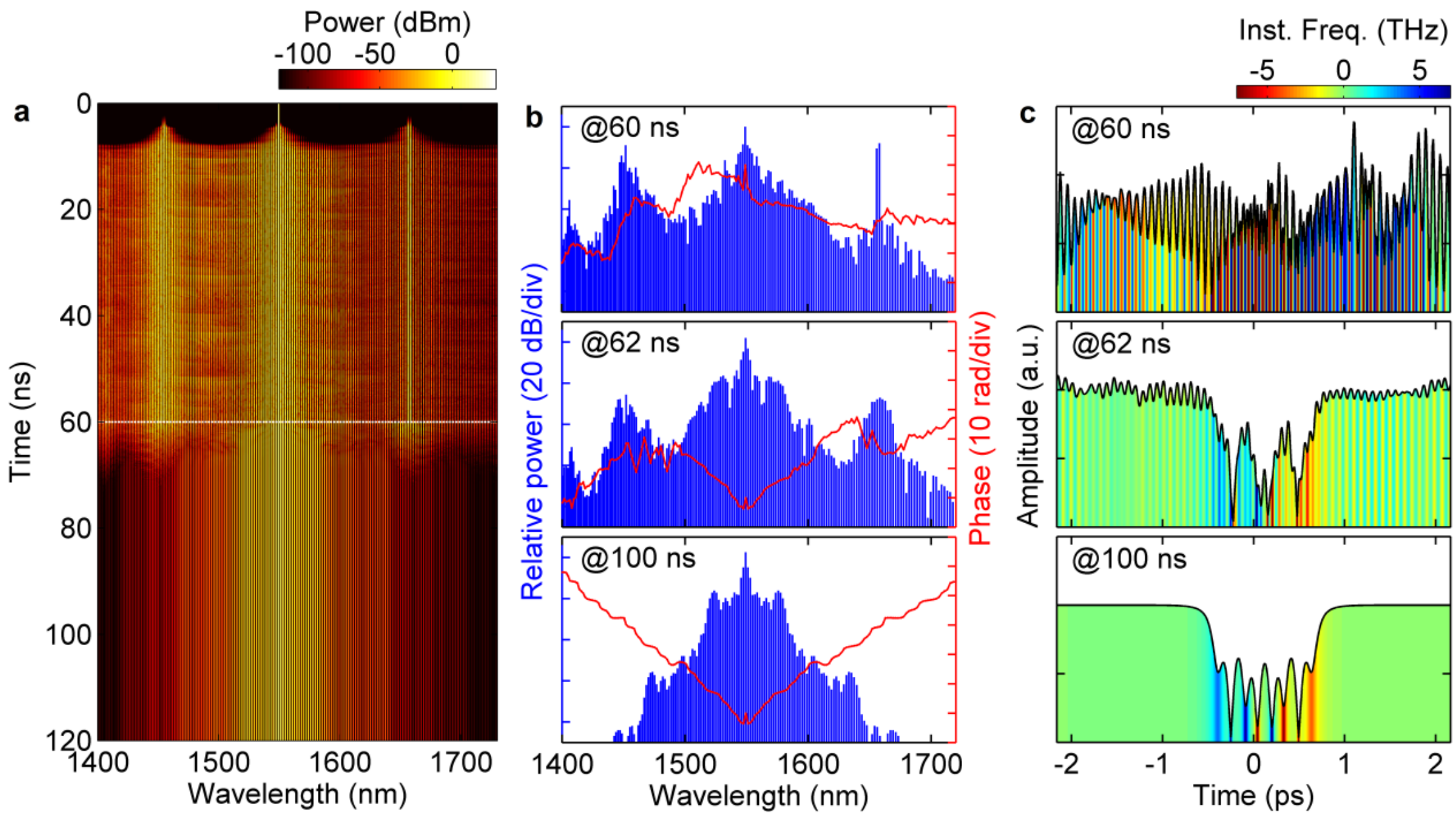

Fig. S10 | Numerical simulation for the microring in Fig. 1a of the main paper. **a**, Evolution of the comb spectrum versus the slow time. **b & c,** Transient comb spectra and waveforms at different slow times. Since the general L-L equation uses an $\exp(-i\omega t)$ convention [S1]-[S2] while our experimental measurements use the $\exp(i\omega t)$ convention common in ultrafast optics [S14], the comb phase from the simulation is adjusted to be consistent with the $\exp(i\omega t)$ convention in order to facilitate comparison of subplots b with the experiments.

## 6. Fronts and dark solitons

In simulations, we found that the formation of dark solitons and breathers is related to interactions of fronts which connect the two stable steady-state solutions in the cavity [S10]-[S12]. Fronts can be formed from weak perturbations in an externally driven Kerr resonator subject to normal dispersion [S13]. To show the physics more clearly, we did some simulations using the normalized L-L equation. We found that dark solitons with different features can be excited by different initial intracavity fields even when the driving amplitude and the phase detuning are the same. Figure S11 shows the results of one simulation example. The phase detuning $\Delta = 5 > \sqrt{3}$ so that bistable steady-state intracavity solutions exist (Fig. S11a). The initial field is a square dark pulse which contains two edges connecting the upper and lower branches (Fig. S11b). The amplitude and phase of the dark pulse top are equal to the amplitude and phase of the upper branch, while the amplitude and phase of the bottom are equal to those of the lower branch. The distance between the two edges (i.e., the width of the initial dark square pulse) is 6. The stable structure evolved from the initial field is shown in Figs. S11c & S11d. Figures S11e & S11f show another possible stable solution which is evolved from a slightly narrower square dark pulse (of which the initial width is 4). Breathers which are oscillating dark solitons in slow time are also possible. Figures S11g & S11h show a breather which is excited by a square dark pulse with a width of 2, and then transitions to a stationary dark soliton when the phase detuning is slightly increased (Fig. S11g) or the driving amplitude is slightly reduced (Fig. S11h). This transition is similar to that observed in our experiments (see Fig. 1d and Fig. 3d of the main paper).

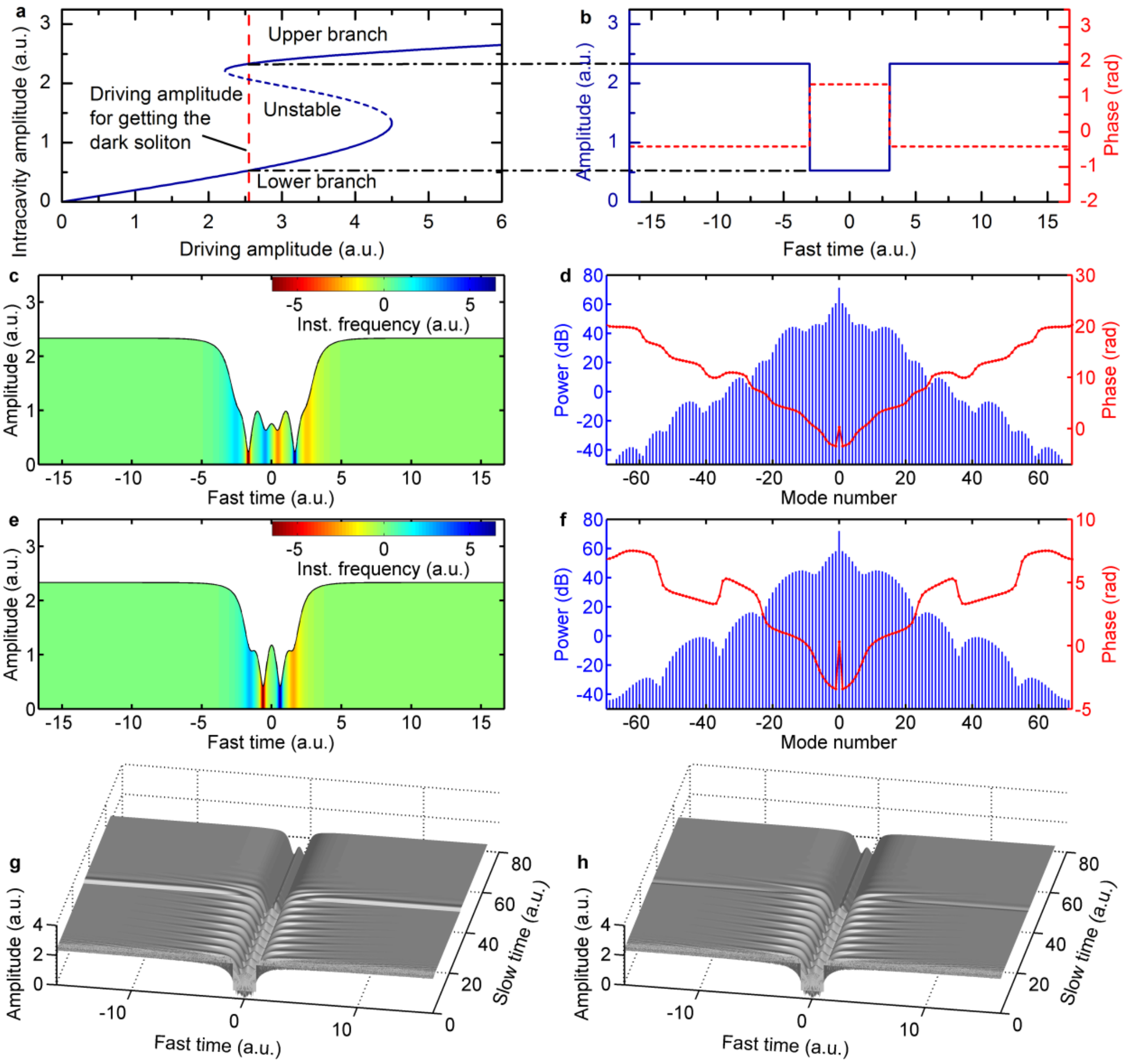

Figure S11 | Numerical simulation of fronts and dark solitons in normal dispersion region. **a**, Steady-state intracavity amplitude versus the external driving amplitude. **b**, Initial intracavity field which is a square dark pulse. The width is 6. The phase detuning $\Delta = 5$. The driving amplitude $S = 2.55$. The top amplitude and phase are equal to the upper-branch steady-state values, while the bottom amplitude and phase are equal to the lower-branch steady-state values. **c**-**d**, Stable dark soliton evolved from the initial field shown in subplot b. **c**, Time-domain amplitude and phase. The color shows frequency chirp. **d**, Frequency-domain comb amplitude and phase. **e**-**f**, Another possible solution of dark soliton with different features compared to subplots c & d. The soliton is excited by an initial square dark pulse with a width of 4. **e**, Time-domain amplitude and phase. **f**, Frequency-domain comb amplitude and phase. **g**-**h**, Breather which is excited by an initial square dark pulse with a width of 2. The breather transitions to a stable dark soliton after (**g**) the phase detuning is increased to 5.5 or (**h**) the driving amplitude is reduced to 2.45 at slow time 33. As in the previous figure, the phase plotted in subplots b, d, f is adjusted to

conform to the $\exp(i\omega t)$ convention used in ultrafast optics, thereby facilitating comparison with our experimental plots.

**References for supplementary information**